\documentclass[prx,aps,twocolumn,reprint,amsmath,floatfix,amssymb,showpacs,superscriptaddress]{revtex4-1}
\usepackage{xcolor}
\usepackage{subfigure}
\usepackage{pict2e}
\usepackage{multirow}
\newcommand{\be}{\begin{equation}}
\newcommand{\ee}{\end{equation}}
\newcommand{\bea}{\begin{eqnarray}}
\newcommand{\eea}{\end{eqnarray}}
\newcommand{\sbe}{\small\begin{equation}}
\newcommand{\see}{\end{equation}\normalsize}
\newcommand{\sbea}{\small\begin{eqnarray}}
\newcommand{\seea}{\end{eqnarray}\normalsize}
\newcommand{\fig}[1]{Fig.~\ref{#1}} 
\newcommand{\XXZ}{XXZ$\text{-}J_{1}\text{-}J_{3}$ }
\newcommand{\JKG}{$JK\Gamma\Gamma^{\prime}$ }

\usepackage{graphicx}
\usepackage{dcolumn}
\usepackage{bm}
\usepackage{color}

\usepackage{epstopdf}

\usepackage{amsmath, amssymb, graphics, setspace}

\newcommand{\mathsym}[1]{{}}
\newcommand{\unicode}[1]{{}}

\usepackage{titlesec}
\usepackage{bm}
\usepackage{comment}
\usepackage{verbatim}

\usepackage{graphicx}
\usepackage{subfigure}
\usepackage{tabularx}
\usepackage{empheq}
\usepackage{physics}

\usepackage{color}
\usepackage[colorlinks,bookmarks=false,citecolor=blue,linkcolor=red,urlcolor=blue]{hyperref}

\definecolor{darkred}{rgb}{0.7,0.0,0.0}

\definecolor{darkblue}{rgb}{0,0.02,0.45}

\definecolor{darkgreen}{rgb}{0.02,0.45,0.0}

\definecolor{violet}{rgb}{0.8,0.2,0.6}

\def\be{\begin{equation}}
\def\ee{\end{equation}}
\def\bea{\begin{eqnarray}}
\def\eea{\end{eqnarray}}

\def\vec{\mathbf}

\begin{document}

\preprint{APS/123-QED}

\title{Geometrical frustration versus Kitaev interactions in BaCo$_2$(AsO$_4$)$_2$}
\author{Thomas Halloran}
\thanks{These authors contributed equally to this work.}
\affiliation{Institute for Quantum Matter and Department of Physics and Astronomy, Johns Hopkins University, Baltimore MD 21218, USA}
\author{F\'elix Desrochers}
\thanks{These authors contributed equally to this work.}
\affiliation{%
 Department of Physics, University of Toronto, Toronto, Ontario M5S 1A7, Canada
}%

\author{Emily Z. Zhang}
\thanks{These authors contributed equally to this work.}
\affiliation{%
 Department of Physics, University of Toronto, Toronto, Ontario M5S 1A7, Canada
}%

\author{Tong Chen}
\affiliation{Institute for Quantum Matter and Department of Physics and Astronomy, Johns Hopkins University, Baltimore MD 21218, USA}
\author{Li Ern Chern}
\affiliation{%
 T.C.M. Group, Cavendish Laboratory, University of Cambridge, Cambridge CB3 0HE, United Kingdom
}%
\author{Zhijun Xu}
\affiliation{NIST Center for Neutron Research, Gaithersburg, Maryland\ 20899, USA}
\affiliation{Department of Materials Science and Engineering, University of Maryland, College Park, Maryland 20742, USA}
\author{Barry Winn}
\affiliation{Neutron Scattering Division, Oak\ Ridge\ National\ Laboratory,\ Oak\ Ridge,\ Tennessee\ 37831, USA}
\author{M. K. Graves-Brook}
\affiliation{Neutron Scattering Division, Oak\ Ridge\ National\ Laboratory,\ Oak\ Ridge,\ Tennessee\ 37831, USA}
\author{M. B. Stone}
\affiliation{Neutron Scattering Division, Oak\ Ridge\ National\ Laboratory,\ Oak\ Ridge,\ Tennessee\ 37831, USA}
\author{Alexander I. Kolesnikov}
\affiliation{Neutron Scattering Division, Oak\ Ridge\ National\ Laboratory,\ Oak\ Ridge,\ Tennessee\ 37831, USA}
\author{Yiming Qiu}
\affiliation{Neutron Scattering Division, Oak\ Ridge\ National\ Laboratory,\ Oak\ Ridge,\ Tennessee\ 37831, USA}
\author{Ruidan Zhong}
\email{rzhong@sjtu.edu.cn}
\affiliation{Department of Chemistry, Princeton University, Princeton, NJ 08544, USA}
\affiliation{Tsung-Dao Lee Institute and School of Physics and Astronomy,
Shanghai Jiao Tong University, Shanghai 200240, China}
\author{Robert Cava}
\affiliation{Department of Chemistry, Princeton University, Princeton, NJ 08544, USA}
\author{Yong Baek Kim}
\email{ybkim@physics.utoronto.ca}
\affiliation{%
 Department of Physics, University of Toronto, Toronto, Ontario M5S 1A7, Canada
}

\author{Collin Broholm}
\email{broholm@jhu.edu}
\affiliation{Institute for Quantum Matter and Department of Physics and Astronomy, Johns Hopkins University, Baltimore MD 21218, USA}
\affiliation{NIST Center for Neutron Research, Gaithersburg, Maryland\ 20899, USA}
\affiliation{Department of Materials Science and Engineering,
The\ Johns\ Hopkins\ University, Baltimore, Maryland\ 21218, USA}

\date{\today}

\begin{abstract}
Recently, Co-based honeycomb magnets have been proposed as promising candidate materials to host the Kitaev spin liquid state. One of the front-runners is BaCo$_2$(AsO$_4$)$_2$ (BCAO), where it was suggested that the exchange processes between Co$^{2+}$ ions via the surrounding edge-sharing oxygen octahedra could give rise to bond-dependent Kitaev interactions. In this work, we present and analyze comprehensive inelastic neutron scattering studies of BCAO with fields in the honeycomb plane. Combining the constraints from the magnon excitations in the high-field polarized state and the inelastic spin structure factor measured in zero magnetic field, we examine two leading theoretical models: the Kitaev-type \JKG model and the \XXZ model. We show that the existing experimental data can be consistently accounted for by the \XXZ model but not by the \JKG model, and we discuss the implications of these results for the realization of a spin liquid phase in BCAO and more generally for the realization of the Kitaev model in cobaltates. 
\end{abstract}

\maketitle

\section{\label{sec:Introduction} Introduction}

There have been tremendous efforts to realize the Kitaev spin liquid (KSL), which is the ground state of the exactly solvable spin-1/2 model with the bond-dependent Ising interactions on the honeycomb lattice. 
The KSL is characterized by exotic excitations such as $\mathbb{Z}_2$ gauge fluxes and Majorana fermions \cite{Kitaev2006AnyonsBeyond,hermanns2018physics,takagi2019concept,motome2020materials,balents2010spin,wen2004quantum, savary2016quantum,zhou2017quantum,broholm2020quantum, knolle2019field}, which may serve as a useful platform for topological quantum computations \cite{freedman2003topological, nayak2008non}. In the majority of proposed candidate Kitaev materials, the physical mechanism to induce the bond-dependent Kitaev interaction relies on exchange processes between spins formed by $4d$ and $5d$ ions with large spin-orbit coupling (SOC), such as Ru$^{3+}$ and Ir$^{4+}$ \cite{Jackeli2009MottModels, Rau2016Spin-OrbitMaterials}. However, in such materials, relatively large non-Kitaev interactions induce magnetic order at low temperature in place of the KSL \cite{Rau2014GenericLimit, rau2014trigonal,Xu2013AbsoluteData, Plumb2014-Lattice,Banerjee2017Neutron-RuCl3,Banerjee2018Excitations-RuCl3,Biffin2014UnconventionalDiffraction,Ducatman2018Magnetic-Li2IrO3,Majumder2018Breakdown-Li2IrO3, Witczak-Krempa2014CorrelatedRegime,Takayama2015HyperhoneycombMagnetism, gohlke2018quantum, schaffer2012quantum}. One is thus left with the hope of potentially driving a system in close proximity to a KSL ground state using external tuning parameters such as the magnetic field, external pressure, and chemical doping \cite{gordon2019theory,sorensen2021heart, kim2016revealing, Banerjee2018Excitations-RuCl3,Zheng2017Gapless-RuCl3,Baek2017Evidence-RuCl3,Kasahara2018MajoranaLiquid, lee2020magnetic, chern2020magnetic, li2022magnetic}. 

\begin{figure}[t]
    \centering
    \includegraphics[width=0.95\columnwidth]{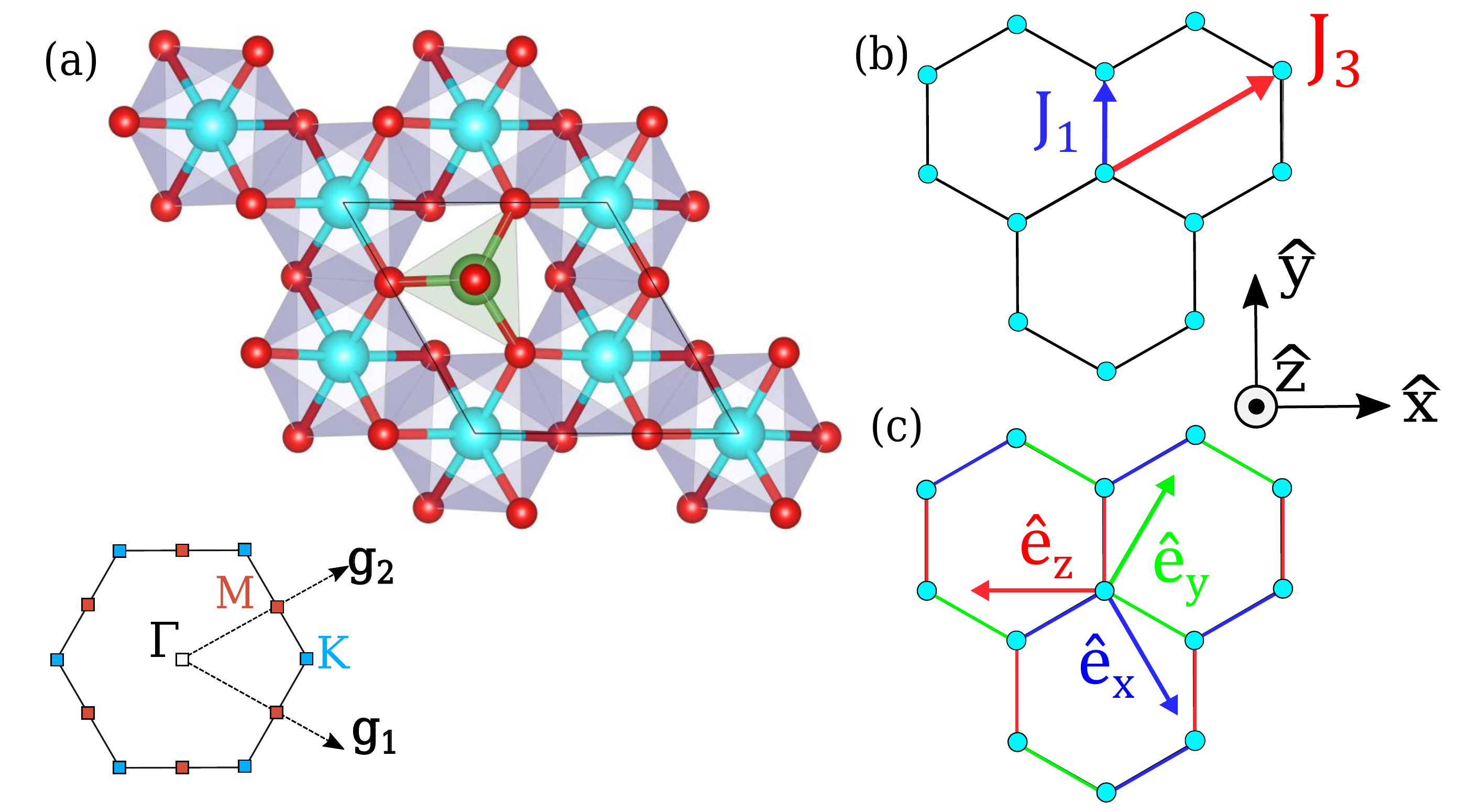}
    \caption{(a) Honeycomb lattice of BCAO viewed along the $c$-axis. Cobalt ions are in blue, oxygen in red, and arsenic in green. The first Brillouin zone of the hexagonal lattice is presented with the reciprocal lattice vectors $\textbf{g}_1$ and $\textbf{g}_2$. The high symmetry points $M$ and $K$ are depicted by the red and blue squares, respectively. (b) Representation of the couplings in the \XXZ model. The crystallographic frame is shown to the right in the black axes. (c) Local spin-frame for Kitaev interaction on the honeycomb lattice as seen from above. The $x$, $y$, and $z$ bonds are shown in blue, green, and red coloring respectively.} \label{fig:structure_fig}
\end{figure}    

A new mechanism for a more ideal material realization of the Kitaev model has recently been put forward for high-spin 3$d^{7}$ cobaltates in the ($S=3/2$, $L=1$) configuration \cite{Liu2018PseudospinModel,Sano2018Kitaev-HeisenbergInsulators}. It was suggested that various exchange processes involving the $t_{2g}$ and $e_{g}$ orbitals can lead to a dominant ferromagnetic Kitaev interaction between the $J_{\text{eff}}=1/2$ pseudospins with relatively weak non-Kitaev couplings. The honeycomb cobaltates are exciting as Kitaev candidate materials from an experimental point of view because large single crystals with few detectable structural defects can be grown, which is not the case for van der Waals materials like $\alpha$-RuCl$_3$ and $\text{H}_3\text{LiIr}_2\text{O}_6$. Additionally, Kitaev materials based on Ir$^{4+}$ are hampered by significant neutron absorption from $^{191}$Ir and a steep reduction of intensity with increasing wave vector transfer $Q$ due to the magnetic form factor, complicating neutron scattering studies. $\text{Co}^{2+}$-based materials face none of these challenges.  

One candidate 3$d^{7}$ Kitaev material is the layered honeycomb magnet BaCo$_2$(AsO$_4$)$_2$ (see Fig.~\ref{fig:structure_fig}(a)), hereafter referred to as BCAO. Heat capacity, magnetization, and neutron scattering studies show that BCAO develops incommensurate antiferromagnetic (AFM) order at $T_N$=5.5~K with an ordering wavevector of $\vec{k}_{c}=(0.27,0,-1.31)$. At low temperatures BCAO undergoes two successive phase transitions with an applied in-plane field: first at $\mu_{0}H_{c1}=0.33$~T into a commensurate ordered state with wavevector $\vec{k}_{c}=(1/3,0,-1.31)$, and a second into a uniform and almost fully magnetized state at a critical field $\mu_{0}H_{c2}=0.55$~T \cite{Zhong2020Weak-fieldHoneycomb, Regnault1977MagneticBaCo2AsO42,Regnault1990PhaseMagnets, Regnault2018Polarized-neutronMagnet}. The critical field and temperature are very small for a transition metal oxide, indicating a tenuous nature of the incommensurate magnetic order. Yet the field-driven state is almost fully magnetized, which indicates BCAO has dominant ferromagnetic interactions. Here we report measurements of the spin-wave excitations in the field-driven ferromagnetic (FM) state and use these in conjunction with the other thermomagnetic data to establish the underlying spin Hamiltonian for BCAO.

Notwithstanding the suggestions that Kitaev interactions might be realized in BCAO, an ab-initio study suggested an entirely different description \cite{Das2021XYCobaltates}. This investigation reported that trigonal distortions play an important role in the physics of BCAO and that it forms an easy-plane XXZ magnet with significant geometrical frustration associated with large third nearest-neighbor AFM interactions (\XXZ model). Previous studies on isostructural BaCo$_2$(PO$_4$)$_2$ were able to completely describe the ground state using a very similar model \cite{Nair2018Short-rangeBaCo2PO42}. 

In this work, we scrutinize the magnetic interactions in BCAO via a combination of inelastic neutron scattering studies (INS) and theoretical analyses. Little experimental work has been done to distinguish the \JKG model with large Kitaev interactions from the \XXZ model in BCAO or, in fact, any cobaltates. Here we report two detailed INS experiments: (1) a zero-field experiment with wavevector transfer {\bf Q} in the $(hk0)$ scattering plane and (2) an experiment with {\bf Q} in the $(h0l)$ plane and a magnetic field applied along the perpendicular [010] direction. This mapping of the magnetic excitation spectrum enables a critical examination of the two competing models. Our results are summarized as follows: We tightly constrain where BCAO may lie in parameter space for each model using linear spin-wave theory in the field-polarized regime. Within these constrained subspaces, we then examine the magnetic ground states that are realized classically. While both models accommodate incommensurate magnetic orders, the relevant spiral order with an ordering wavevector between the $\Gamma$ and $M$ points can only be stabilized in the \XXZ model. Additionally, molecular dynamics (MD) calculations demonstrate that the dynamical spin-structure factor at zero field seems incompatible with large ferromagnetic Kitaev interaction, whereas all current data can be adequately accounted for using an \XXZ spin Hamiltonian with only weak bond-dependent interactions. Finally, for a suitably chosen parameter set of the \XXZ model, we are able to account for our measured in-plane magnetization curve, which features three successive plateaus separated by two field-induced phase transitions and only weak anisotropy for different in-plane directions. This is not the case for the \JKG model. We thus find compelling evidence indicating that BCAO is not a realization of the Kitaev model but rather a geometrically frustrated easy-plane magnet with predominantly isotropic in-plane exchange interactions. We further provide a highly constrained set of coupling constants for this model. 

Our work highlights the need for a thorough examination of the exchange interactions in the honeycomb cobaltates, a number of which have been proposed as KSL candidates. While a set of exchange parameters with a large Kitaev interaction may reproduce the observed INS excitation spectra in the high-field polarized state with a linear spin-wave theory (LSWT) fit, it is important to examine whether this parameter set is consistent with the zero-field ordered state and thermodynamic measurements. Even though BCAO does not appear to have strong Kitaev interactions based on our work, it still may be close to a QSL driven by geometrical frustration and the quasi-2D nature of the spin system. The emergence of a continuum for an out-of-plane field upon the suppression of order revealed in THz spectroscopy \cite{Zhang2021In-BaCo_2AsO_4_2} might be a signature of such a proximate QSL whose characteristics would most likely differ significantly from the non-abelian KSL. As a result, we speculate on a potential out-of-plane field-induced QSL in BCAO using the extracted exchange parameters and propose directions for future works.


\begin{figure*}
    \centering
    \includegraphics[width=1.0\columnwidth]{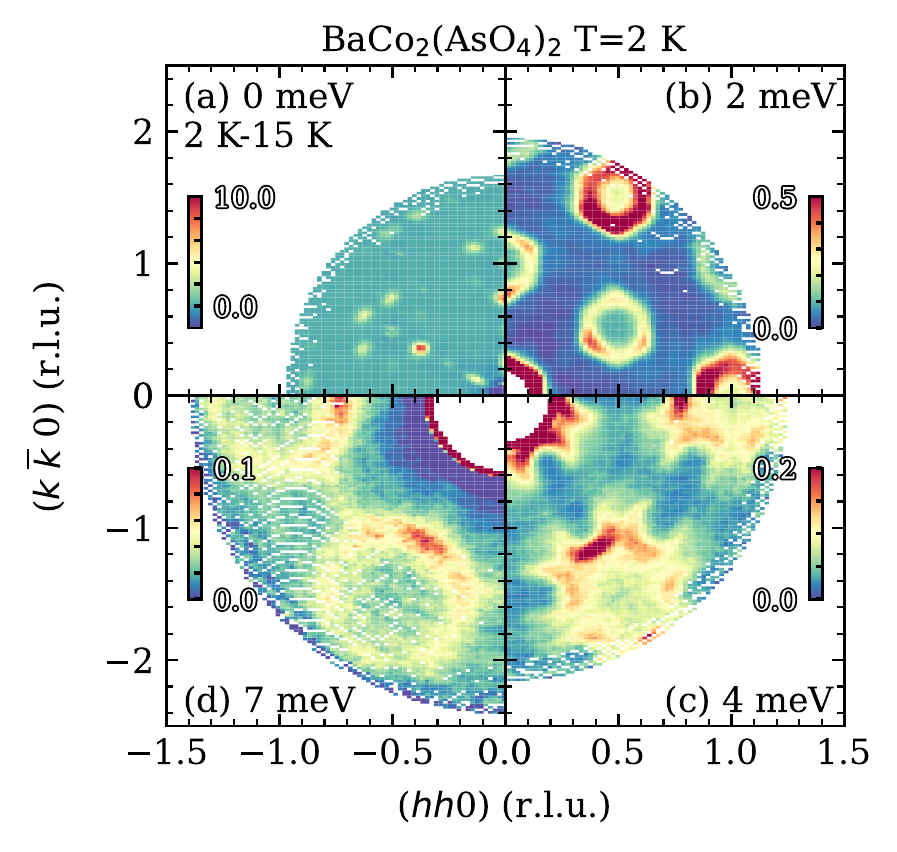}
    \includegraphics[width=1.0\columnwidth]{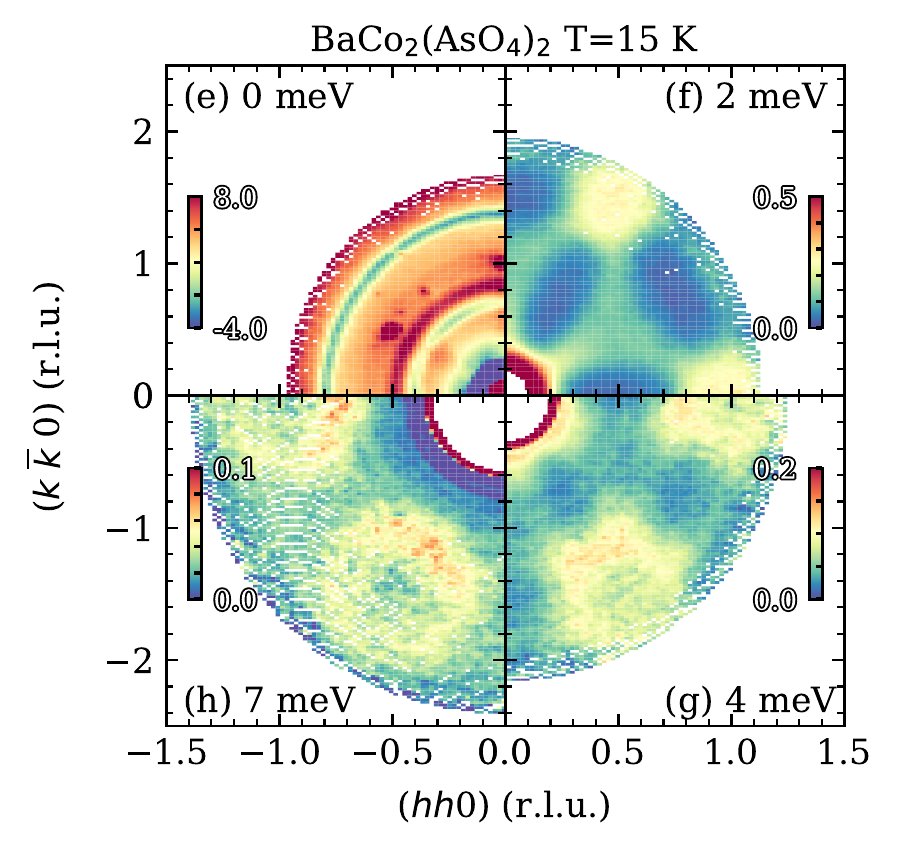}
    \caption{Zero field neutron scattering from BaCo$_2$(AsO$_4$)$_2$ as a function of energy transfer and temperature in the $(hk0)$ scattering plane. Panels (a-d) show scattering at 2 K, and (e-h) at 15 K. Panel (e) has been used as a background to panel (a) to make the magnetic diffraction from the ordered state more visible. Energy transfers for each slice are as labeled. Sample out measurements were used as backgrounds for all measurements.}
    \label{fig:macs15K_inel_fig}
\end{figure*}

\section{\label{sec:Experimental Methods}Experimental Methods}

\subsection{\label{sec:Materials Synthesis}Materials Synthesis}
Two sets of crystals were grown for these experiments. The MACS experiment totaled 0.88(1) g of sample, and the SEQUOIA and HYSPEC experiments totaled 0.96(1) g. All were grown by a flux method and have a dark purple coloring \cite{Zhong2020Weak-fieldHoneycomb}. No stacking faults or secondary phases were detected through single-crystal x-ray diffraction. The $T=293(2)$ K lattice constants observed are $a=b=5.007(1)$~\AA, and $c=23.491(5)$~\AA~ \cite{Dordevic2008BaCo2AsO42}, and the system crystallizes in space group $R\bar{3}$ (No. 148). 

\subsection{\label{sec:Magnetic Neutron Methods}Magnetic Neutron Scattering}
Three neutron scattering measurements were performed. The first was performed on the MACS instrument at NIST in an "orange" $^4$He flow cryostat with a 1.5 K base temperature. For this experiment, many plate-like crystals of BCAO were coaligned in the $(hk0)$ scattering plane. The final neutron energy setting was fixed at E$_f$=5~meV, and the monochromator was in the double-focusing high flux mode. Measurements were performed at $T=1.7(1)$~K and $T=15.0(1)$~K, with 16 hours and 18 hours of counting time respectively. Measurements were taken at energy transfers of 0~meV, 2~meV, 4~meV, and 8~meV for both temperatures. No magnetic field was applied in this measurement. 

The second measurement was taken at the SEQUOIA instrument at ORNL on a second set of crystals with a total mass of 0.96(1) g. These crystals were coaligned in the $(h0l)$ scattering plane and the experiment was performed in a vertical field magnet such that the field was applied along the b-axis, which is parallel to the $(\bar{1}20)$ direction. Measurements were taken using the $E_i=21$~meV high flux configuration with Fermi chopper frequency of 120 Hz. Measurements were conducted with the samples at $T=2$~K, 15~K, and 50~K and total proton charge of 74~C, 80~C, and 40~C, respectively. The resulting data are displayed in the supplementary information. 

The third experiments was run on the HYSPEC instrument at ORNL using the same set of crystals as on SEQUOIA aligned in the $(h0l)$ plane. We used an Oxford instruments 14~T vertical field magnet with HYSPEC in the unpolarized high flux 300~Hz configuration with incident energies $E_i=6$~meV and $E_i=27$~meV. The main configuration was the low energy $E_i=6$~meV mode at fields of 0~T, 0.4~T, 0.55~T, 0.75~T, 1~T, 2~T, 3~T, 4~T, and 5~T. The net proton charge on target for each field was 140~C, 72~C, 30~C, 101~C, 30~C, 30~C, 75~C, 30~C, and 30~C respectively, with a Fermi chopper frequency of 360 Hz. The higher energy $E_i=27$~meV mode was used for $\mu_{0}H=3$~T with a total proton charge of 44~C and a 420~Hz Fermi chopper frequency. Analysis of the scattering from ORNL experiments was performed using the MANTID software package \cite{Arnold2014MantidExperiments}. 
\begin{figure*}
    \centering
    \includegraphics[width=1.0\textwidth]{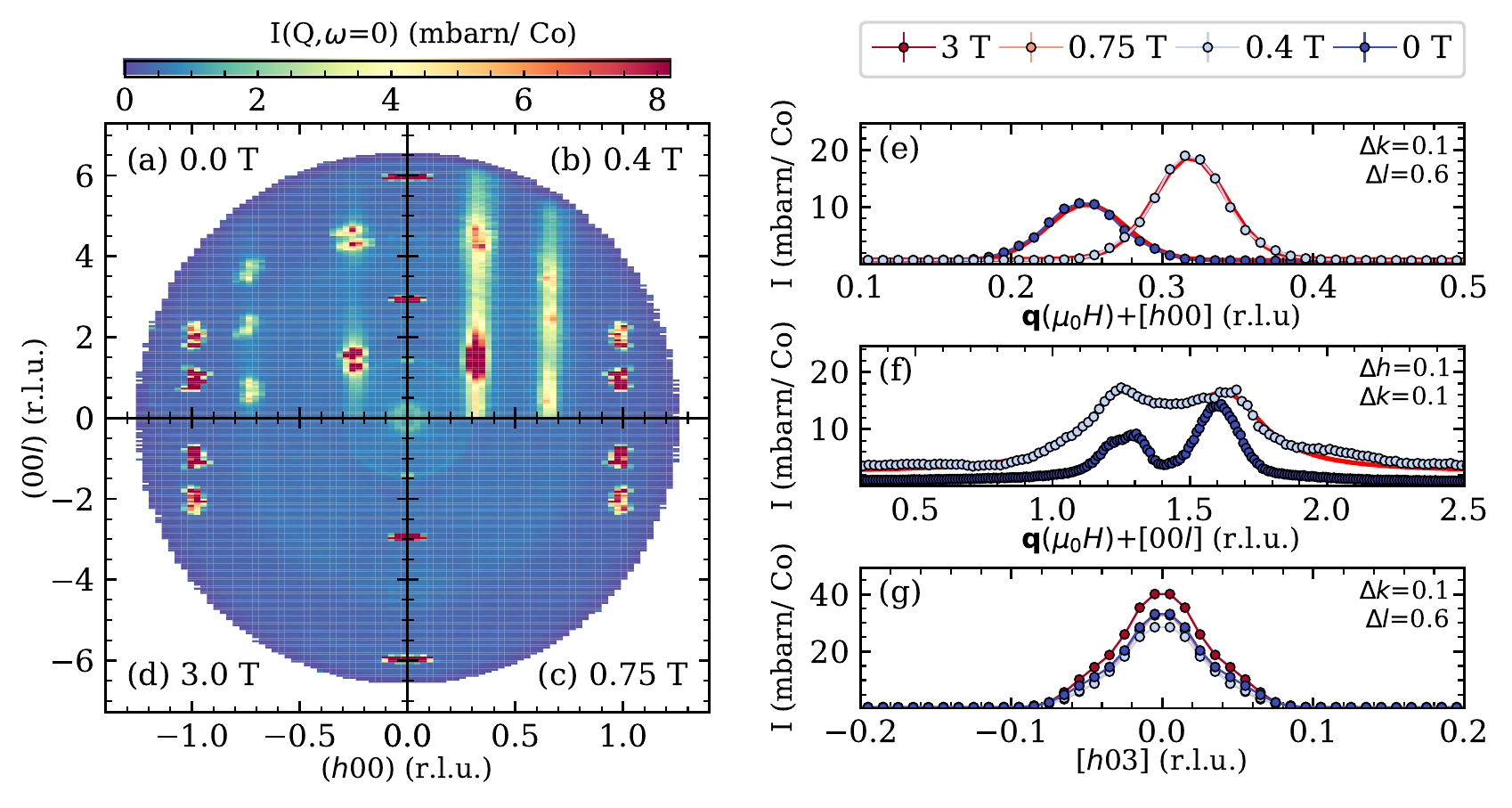}
    \caption{Elastic magnetic scattering from BaCo$_2$(AsO$_4$)$_2$ at $T=1.7$ K (a-d) as a function of field applied perpendicular to the scattering plane and along the $(\bar{1}20)$ direction. The scattering has been averaged in the energy transfer window of $\pm$0.1~meV, and $k\in[-0.1,0.1]$ along the $(0k0)$ direction. The short ranged nature of the correlations is evident at zero field (a) and in the commensurate ordered state at $\mu_{0}H=0.4$~T (b) through the broadness of the peaks in the $(00l)$ direction. Panels (e-f) further highlight this, comparing cuts at the respective incommensurate wavevectors $\textbf{q}(H) \equiv \{ q_h (H), 0, q_l (H) \} $. For $\mu_{0}H=0$ T, $\textbf{q}=(0.27,0,-1.31)$ and for $\mu_{0}H=0.4$~T $\textbf{q}=(1/3,0,-1.31)$. Panel (g) compares cuts along the $(h00)$ direction over the $(003)$ structural Bragg peak. The increase in intensity in the field-polarized 3 T phase may be directly accounted for by the induced in-plane moment. The full width of the averaging windows for cuts in (e-g) are shown in the top right corner of each panel. }
    \label{fig:hys_el_fig}
\end{figure*}

\begin{figure}
    \centering
    \includegraphics[width=1.0\columnwidth]{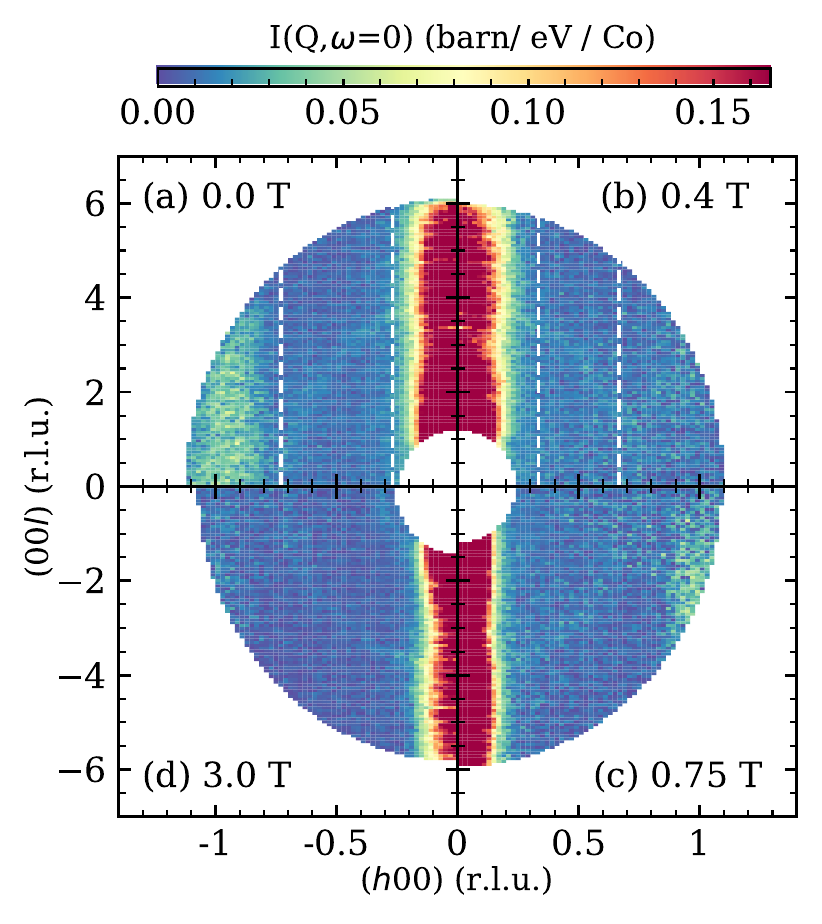}
    \caption{Constant energy slices around the $\Gamma$ point gap energy at four representative fields. The energy windows for panels (a-d) begin at $\hbar\omega=0.4$~meV and extend to $\hbar\omega=1.58$~meV, 1.80~meV, 1.90~meV, and 2.29~meV, respectively. Dashed white lines are shown at the incommensurate ordering wavevector for each respective field, with no lines for the magnetized phases without AFM order.}
    \label{fig:hys_h0l_gap}
\end{figure}
\subsection{In-plane Magnetization}
\begin{figure*}
    \centering
    \includegraphics[width=1.0\textwidth]{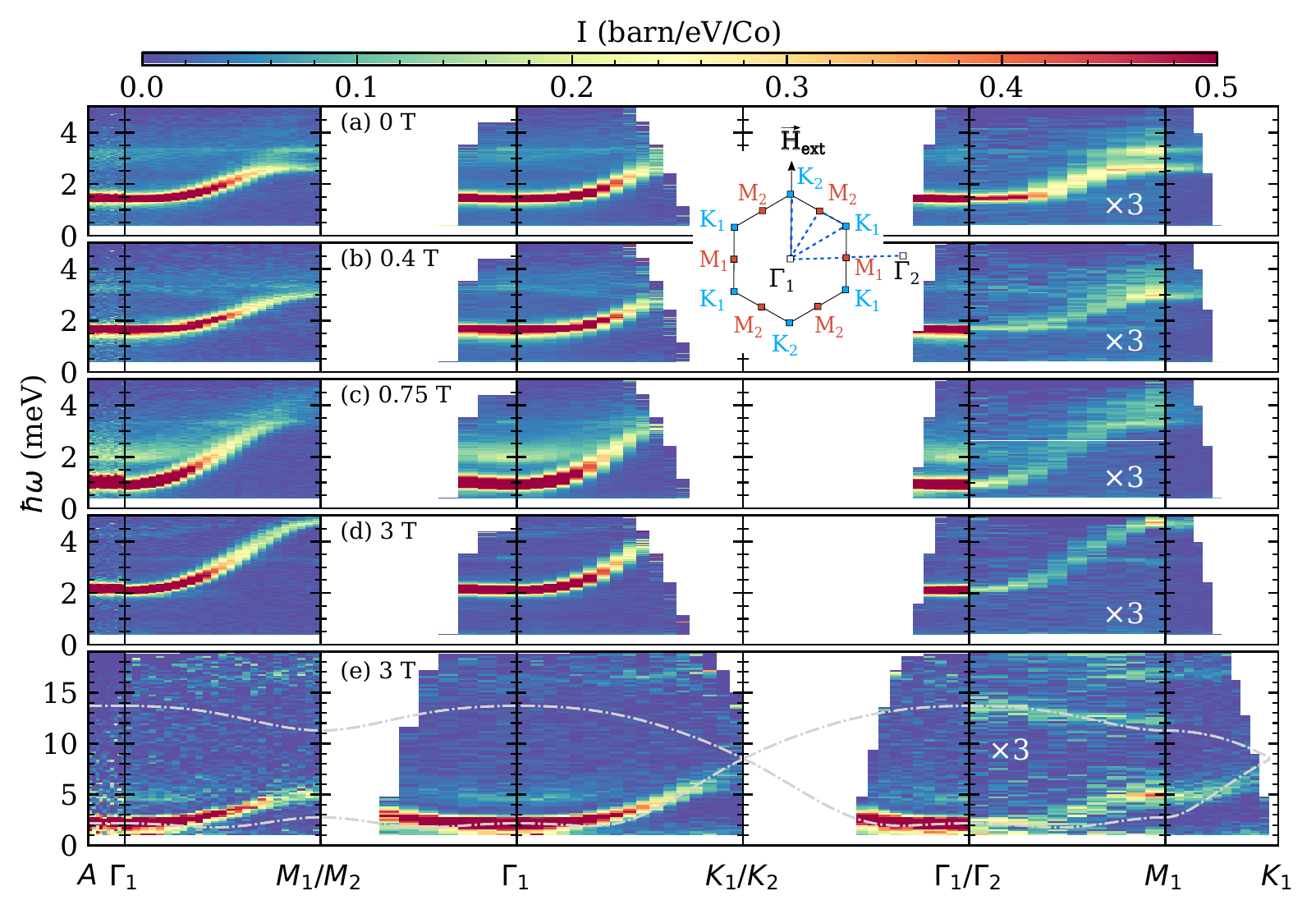}
    \caption{Magnetic excitation spectrum of BCAO as a function of field along high symmetry directions in the hexagonal Brillouin zone at $T$=1.7 K. The path in r.l.u. notation is $(00\left[\ell +\frac{1}{2}\right])$-$(00\ell)$-$(\frac{1}{2}0\ell)/(0\frac{1}{2}\ell)$-$(00\ell)$-$(\frac{1}{3}\frac{1}{3}\ell)/(\frac{\bar{1}}{3}\frac{2}{3}\ell)$-$(00\ell)/(10\ell)$-$(\frac{1}{2}0\ell)$-$(\frac{1}{3}\frac{1}{3}\ell)$. This is represented by the dashed blue lines pictured in the diagram of the Brillouin zone, where $\vec{H}_{\text{ext}}$ denotes the direction of the applied field. In (a-d) $\ell\in[1,3]$ for the in-plane slices in the high resolution $E_i$=6~meV configuration. For the $\Gamma$-$A$ direction, the path of $L$ to $L+\frac{1}{2}$ is averaged over several Brillouin zones ($L=1,2,3,4$). For the $\Gamma_2-M_1$ path, intensity has been scaled by a factor of three as indicated and $R\bar{3}m$ symmetry operations have been applied to enhance statistics. The transverse-$\textbf{Q}$ averaging window is 0.1 \AA$^{-1}$ in all cases. Panel (e) shows the $\mu_{0}H=3$~T and $E_i$=27~meV high energy configuration where $\ell\in [1,9]$. Data dominated by the tail of elastic scattering at energies below the FWHM elastic instrumental resolution of $\Delta\hbar\omega$=0.15~meV for the low energy configuration and $\Delta\hbar\omega=$1.5~meV for the high energy configuration have been masked.} \label{fig:hys_inel_fig}
\end{figure*}
Low $T$ Magnetization measurements were performed on a high-quality single crystal of BCAO as a function of the applied magnetic field strength and orientation in the basal plane. The sample was a 0.88(2) mg plate that was aligned using a Laue diffractometer to an accuracy better than 1 degree. The sample was mounted on a quartz rod and oriented such that the applied field was along the $\hat{y}$ direction, which is equivalent to the $(1\bar{1}0)$ direction as indicated in Fig.~\ref{fig:structure_fig}(b). The sample was then cooled from 300 K to 2 K in zero field. A full hysteresis loop was measured at $T=2$ K in the applied field range $\mu_0H\in[-1,1]$~T. The sample was then warmed to 20 K, cooled back to $T=2$ K in zero field, and the hysteresis loop was measured again to confirm that the measured curves were reproducible. The sample was then carefully rotated in $15^\circ$ steps until the field was along the $\hat{x}$ direction, repeating the zero field cooled magnetization scans for each orientation. Here we only present data acquired for the $\hat{x}$ and $\hat{y}$ field directions. These measurements were run in a Quantum Design MPMS3 SQUID magnetometer.  

\section{\label{sec:Experimental Results} Experimental Results}

\subsection{Static and dynamic spin correlations}
Constant energy slices in the $(h0\ell)$ scattering plane captured in the MACS experiment are shown in Fig.~\ref{fig:macs15K_inel_fig}. Slices at T=1.7(1) K are shown in Fig.~\ref{fig:macs15K_inel_fig}(a-d), and slices at T=15.0(1) K are in Fig \ref{fig:macs15K_inel_fig}(e-h). Fig.~\ref{fig:macs15K_inel_fig}(a) shows elastic scattering at 1.7 K where the 15 K data in Fig.~\ref{fig:macs15K_inel_fig}(e) was subtracted as a background. Magnetic satellite peaks surrounding the $(0\bar{1}0)$ and $(1\bar{1}0)$ nuclear Bragg peak evidence the incommensurate magnetic order that forms below $T_N$. Magnetic satellite peaks are even apparent near the origin. The different intensities of the six satellite peaks surrounding a given nuclear Bragg peak reflect the structure factor of the magnetic order. The full width at half maximum (FWHM) vertical $Q-$resolution of MACS at ${\bf Q}=(100)$ is $\Delta Q=0.09$~\AA$^{-1}=0.2$c$^*$. Thus this $(hk0)$ plane measurement does not access the nominal ${\bf k}=(0.27,0,-1.3)$ magnetic wave vector. The appearance of the magnetic Bragg peaks in Fig.~\ref{fig:macs15K_inel_fig}(a) despite of this is a first indication of the quasi-2D nature of the magnetic order: The magnetic peaks have tails extending along c$^*$ so they can be detected near $l=0$. This is distinct from resolution effects and is made clearer in the second measurement in the $(h0l)$ scattering plane (Fig.~\ref{fig:hys_el_fig}). With increasing energy transfer a faceted ring of intensity centered at $\Gamma$ points with increasing diameter is observed. This indicates anisotropic dispersive excitations with the lowest energy mode at the $\Gamma$ point as for a ferromagnet.   

Even though BCAO is in the paramagnetic phase at $T=15$~K, dynamic spin correlations give rise to a peak in low energy inelastic neutron scattering at the $\Gamma$ point. This indicates predominantly ferromagnetic correlations and is consistent with previous THz and inelastic neutron spectroscopy studies \cite{Zhang2021In-BaCo_2AsO_4_2,Shi2021MagneticSpectroscopy,Regnault2018Polarized-neutronMagnet}. The spectrum of excitations extends well beyond the thermal scale of $k_BT\approx 1\text{ meV}$. This is characteristic of a quasi-two-dimensional magnet with competing interactions where thermal and quantum fluctuations destabilize long-range magnetic order\cite{PhysRevLett.105.037402}. Hexagonally faceted rings of inelastic magnetic scattering around the $\Gamma$ point can be seen in Fig.~\ref{fig:macs15K_inel_fig}(g) and (h). These data are qualitatively similar to the corresponding low-temperature data in the ordered phase shown in Fig.~\ref{fig:macs15K_inel_fig}(c)-(d), which indicates that dynamic correlations within individual honeycomb planes are established well above the ordering temperature. 
\begin{figure}
     \centering
    \includegraphics[width=1.0\columnwidth]{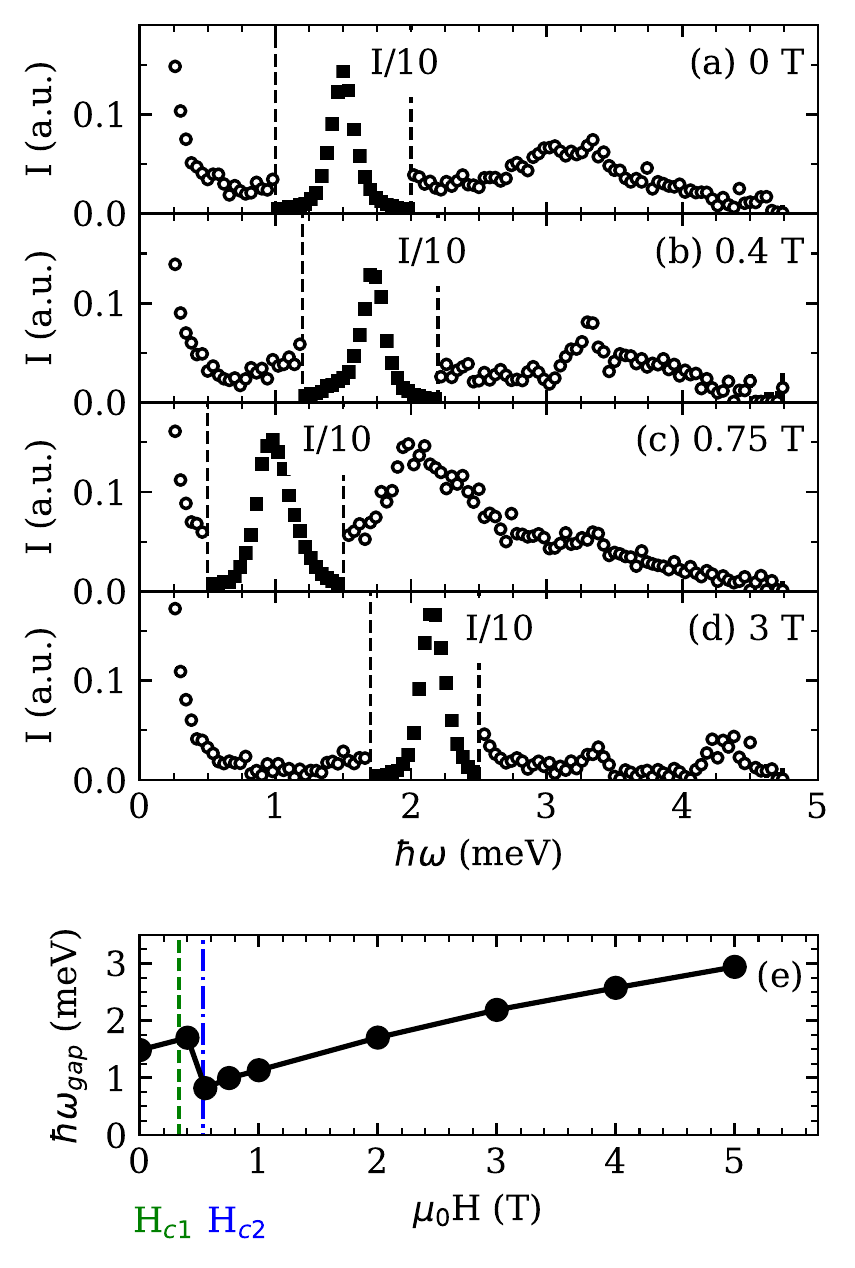}
    \caption{(a-d) Magnetic excitation spectrum for $\rm BaCo_2(AsO_4)_2$ at the $\Gamma$ point for four different magnetic field strengths applied along the $(\bar{1}20)$ direction. Intensity data near the main peak in each spectrum has been scaled down by a factor of 10 and are shown as filled symbols. (e) Gap at the $\Gamma$ point as a function of the magnetic field strength along $(\bar{1}20)$.}
    \label{fig:hys_inel_fig2}
\end{figure}

Elastic scattering in the $(h0\ell)$ plane is shown in Fig.~\ref{fig:hys_el_fig} for three different magnetic fields applied within the honeycomb plane. Complex peak structures are observed at the nuclear Bragg peaks (101) and (102). These reflect the mosaic distribution of the multi-crystal sample that defines the transverse momentum resolution within the $(h0\ell)$ scattering plane. Fig.~\ref{fig:hys_el_fig}(b) and (d) show that elastic scattering beyond the nuclear allowed Bragg peaks is absent for in-plane fields of 0.75~T and beyond. In zero field (Fig.~\ref{fig:hys_el_fig}(a)), magnetic Bragg peaks are seen at the incommensurate ordering wavevector ${\bf k}_{c}=(0.27(1),0,-1.31(1))$. The ordering wavevector becomes commensurate within the honeycomb plane ${\bf k}_{c}=(1/3,0,-1.31)$ in a field of $\mu_0 H=0.4$~T applied along the $(1\bar{2}0)$ direction (Fig.~\ref{fig:hys_el_fig}(c)). In both the 0~T and 0.4~T measurements, the magnetic scattering is extended in momentum space along the $(00\ell)$ direction, which indicates a reduced correlation length along the $c$-axis. 

Fig.~\ref{fig:hys_el_fig}(f) shows cuts through these elastic scattering data along the $(00\ell)$ direction through the magnetic peaks for both $\mu_0H=0$~T and $\mu_0H=0.4$~T. The Lorentzian fits (solid red lines) indicate correlation lengths of $\xi=70(2)$~\AA\ and $\xi=22(1)$~\AA\ along $c$ in the incommensurate and the commensurate phases, respectively. These may be compared to the interlayer spacing of $d=7.64$~\AA, indicating a quasi-two-dimensional order and weak interactions between honeycomb layers. Correlations along $(h00)$, in contrast, are limited by the instrumental resolution from which we infer a correlation length exceeding $\sim$300 \AA\ or $\sim$20$a$ as shown in Fig~\ref{fig:hys_el_fig}(e). This value is found assuming an instrumental resolution equal to the width of the $(003)$ Bragg peak. Fig.~\ref{fig:hys_el_fig}(g) shows that in the fully field polarized state at $\mu_0H$=3 T, the (003) Bragg peak gains strength, which is consistent with magnetic diffraction from the magnetized cobalt. The 24(1)\% increase of the (003) peak intensity in the 3~T applied field corresponds to an induced magnetization of $2.4(7)~\mu_B$/Co, which may be compared to the magnetization of $2.9(1)~\mu_B$/Co obtained from magnetization data.

It is interesting to contrast the $\bf Q$-dependence of the elastic magnetic scattering as depicted in Fig.~\ref{fig:hys_el_fig} with that of the low energy inelastic scattering in Fig.~\ref{fig:hys_h0l_gap}. Integrating over energy transfer from $\hbar\omega=0.4$~meV through the $\Gamma$ point excitation energy for each value of the applied field, we find a rod of scattering extending along $\bf c$ and passing through the $\Gamma$ point. This contrast with the $\bf Q$-dependence of the elastic magnetic scattering, which at 0 T and 0.4 T has a finite in-plane component as apparent in Fig.~\ref{fig:hys_el_fig}(a,c) and also indicated by the dashed lines in Fig.~\ref{fig:hys_h0l_gap}(a,c). The dynamic spin correlations in BCAO thus resemble those of a 2D FM at all fields and contain no evidence of a soft mode with a wave vector matching that of the low field AFM order.

\subsection{Dispersive magnetic excitations}
${\bf Q}-\omega$ slices through inelastic neutron scattering data for $\bf Q$ varying along high symmetry directions in the hexagonal Brillouin zone are shown for four values of magnetic field applied along the $(1\bar{2}0)$ direction in Fig.~\ref{fig:hys_inel_fig}. The momentum space labels used on the horizontal axis are defined in the inset. In the almost fully magnetized state at 3~T (Fig.~\ref{fig:hys_inel_fig}(d) and (e)), the scattering qualitatively follows expectations for a 2D easy plane honeycomb ferromagnet. There is a coherent gapped mode with lowest energy at the $\Gamma$ point. A sharp flat two-magnon mode is visible near 4 meV. At the lower fields (Fig.~\ref{fig:hys_inel_fig}(a-c)), there are strong diffuse contributions to the scattering near twice the field dependent gap energy. At zero field in Fig.~\ref{fig:hys_inel_fig}(a) where the magnetic order has an incommensurate modulation within the basal plane there are multiple modes at the $M$ points indicative of a large unit cell. The first panels of these subplots show the dispersion along the $(00\ell)$ direction, averaged around the $\Gamma$ point. There is no observable dispersion along this direction, again pointing toward very weak magnetic interaction between honeycomb layers. 

Once the AFM order is suppressed at $\mu_0H$=0.75 T, a magnon remains with a gap of $\hbar\omega$=1.2~meV and a two magnon excitation centered at $\hbar\omega=2.4$~meV. This can be more clearly seen through cuts of the intensity at the $\Gamma$ point as a function of field, as shown in Fig.~\ref{fig:hys_inel_fig2}. An unusual feature of the scattering is a difference in the dispersion relation between high symmetry paths that differ only in their orientation with respect to the applied magnetic field and thus the magnetization. The high symmetry zone boundary points nearest to the horizontal scattering plane are denoted $M_1$ and $K_1$, whereas the points closest to the vertical field direction are denoted $M_2$ and $K_2$ (see sketch in Fig.~\ref{fig:hys_inel_fig}). Though the geometry of the instrument limits experimental access along the latter directions, a flattening of the mode may be observed at all non-zero fields in Fig.~\ref{fig:hys_inel_fig} for the $K_2$-$\Gamma_1$ and $M_2$-$\Gamma_1$ paths that lie along and at 30$^o$ to the field direction respectively when compared to the $K_1$-$\Gamma_1$ and $M_1$-$\Gamma_1$ paths that form larger angles with the field direction. This is direct evidence of spin-orbit coupling and anisotropic magnetic interactions. 

Fig.~\ref{fig:hys_inel_fig}(e) shows a complementary higher energy measurement in the 3 T field-polarized phase. A second magnon mode is observed near $\hbar\omega=12$~meV for wave vector transfer between $\Gamma_2$ and $M_1$. For a fully polarized magnet with two magnetic ions per unit cell, one indeed expects two magnon bands. This upper mode provides additional constraints on the model Hamiltonian. At the $\Gamma$ point, the intensity of the upper magnon mode is zero due to the spin-structure factor, so that this excitation is not visible in Raman and THz optical spectroscopy. The broad excitation visible near $\hbar\omega$=15~meV in Fig.~\ref{fig:hys_inel_fig}(e) is a phonon as confirmed by Raman scattering \cite{Zhang2021In-BaCo_2AsO_4_2}.   

 Fig.~\ref{fig:hys_inel_fig2}(a-d) show the spectrum of inelastic magnetic scattering at the $\Gamma$ point. For the lower fields where BCAO is not fully magnetized, a two-magnon continuum is visible as a broad peak centered at twice the gap mode energy. Fig.~\ref{fig:hys_inel_fig}(a) shows an anomaly in the main magnon dispersion relation as it enters the two-magnon continuum\cite{WOS:000235839500040}. The broad two-magnon peak is particularly strong when the one-magnon gap is smallest at 0.75~T (Fig.~\ref{fig:hys_inel_fig2}(c)) and magnons with energies beyond the two-magnon energy scale acquire significant physical width even though BCAO is uniformly magnetized at 0.75 T. Such interaction effects do not occur for the Heisenberg ferromagnet in a field where single magnons are exact eigenstates. At lower fields, the non-collinear and incommensurate nature of the ground state may play an important role in allowing spin waves to interact with the two-magnon continuum. These effects cannot be captured by the conventional $1/S$ expansion and the associated LSWT. When the system is fully magnetized at 3~T, the gap increases, the intensity of the two-magnon excitation decreases, and it approaches the resolution limit (Fig.~\ref{fig:hys_inel_fig2}(d)). Though the magnon dispersion relation still intersects the two-magnon mode, there is no longer an anomaly. LSWT should provide a good account of the one magnon branch in this almost fully magnetized state. In this work, LSWT dispersions were calculated using both analytical methods and the SpinW package \cite{Toth2015LinearStructures}. The full field dependence of the gap at the $\Gamma$ point is shown in Fig.~\ref{fig:hys_inel_fig2}(e) and is consistent with previous THz studies \cite{Zhang2021In-BaCo_2AsO_4_2}.

\section{\label{sec:Analysis} Analysis}

\subsection{\label{subsec: Minimal spin models}Model Spin Hamiltonians}

There are currently two competing theoretical proposals for the microscopic description of BCAO. Following theoretical predictions that a dominant Kitaev interaction can be obtained in high-spin 3$d^{7}$ cobaltates \cite{Liu2018PseudospinModel,Sano2018Kitaev-HeisenbergInsulators}, recent experiments were interpreted in terms of a \JKG model
\begin{align}
    \mathcal{H}_{JK\Gamma\Gamma'} =&  \sum_{\left<i,j\right>\in \gamma} \mathbf{S}_{i}^{T} H_{K,\gamma}^{(1)} \mathbf{S}_{j}, \label{eq:ham_jkggp}
\end{align}
where
\begin{align}
    H_{K,x}^{(1)} &= 
    \begin{pmatrix}
        J+K & \Gamma^{\prime} & \Gamma^{\prime} \\
        \Gamma^{\prime} & J & \Gamma \\
        \Gamma^{\prime} & \Gamma & J
    \end{pmatrix}, H_{K,y}^{(1)} = 
    \begin{pmatrix}
        J & \Gamma^{\prime} & \Gamma \\
        \Gamma^{\prime} & J+K & \Gamma^{\prime} \\
        \Gamma & \Gamma^{\prime} & J
    \end{pmatrix},\nonumber \\
    H_{K,z}^{(1)} &= 
    \begin{pmatrix}
        J & \Gamma & \Gamma^{\prime} \\
        \Gamma & J & \Gamma^{\prime} \\
        \Gamma^{\prime} & \Gamma^{\prime} & J+K
    \end{pmatrix}
    \label{eq:JKGGp_matrix}
\end{align}
with a large ferromagnetic Kitaev coupling and small isotropic and off-diagonal terms (i.e., $K<0$ and $|K|\gg |J|, |\Gamma|, |\Gamma'|$). The spins in this model are defined with respect to the Kitaev frame (KF) defined in Fig.~\ref{fig:structure_fig}(c). The most general form of $H_{K,\alpha}^{(1)}$ consistent with the symmetry of BCAO is given by Eq.~\eqref{eq: general coupling matrix LKF}. The simpler form used here neglects the lifting of $C_{2v}$ symmetry associated with the puckering of the honeycomb layer.

Alternatively, recent \emph{ab initio} calculations \cite{Das2021XYCobaltates} suggest that BCAO can be described by an \XXZ model where the spin Hamiltonian is approximately isotropic within the basal plane
\begin{align}
    \mathcal{H}_{\text{XXZ}\text{-}J_{1}\text{-}J_{3}} =&  \sum_{\langle i,j\rangle\in \gamma} \mathbf{S}_{i}^{T} H_{\text{XXZ},\gamma}^{(1)} \mathbf{S}_{j} + \sum_{\langle\langle\langle i,j\rangle\rangle\rangle} \mathbf{S}_{i}^{T} H_{\text{XXZ}}^{(3)} \mathbf{S}_{j}, \label{eq:ham_j1j3}
\end{align}
with
\begin{align} \label{eq: coupling matrices XXZ-J1-J3}
    H_{\text{XXZ},z}^{(1)} &=  
    \begin{pmatrix}
        J_{xy}^{(1)} + D & E & 0 \\
        E & J_{xy}^{(1)} - D & 0\\
        0 & 0 & J_{z}^{(1)}
    \end{pmatrix},  
    \nonumber \\
    H_{\text{XXZ},y}^{(1)} &= U_{2\pi/3}^T H_{\text{XXZ},z}^{(1)} U_{2\pi/3},\nonumber\\
    H_{\text{XXZ},x}^{(1)} &= U_{2\pi/3} H_{\text{XXZ},z}^{(1)} U_{2\pi/3}^T, \nonumber \\
    H_{\text{XXZ}}^{(3)} &= 
    \begin{pmatrix}
        J_{xy}^{(3)} & 0 & 0 \\
        0 & J_{xy}^{(3)} & 0\\
        0 & 0 & J_{z}^{(3)}
    \end{pmatrix},
\end{align}
where the second sum is taken over third nearest-neighbors, the spins are defined with respect to the  crystallographic frame (CF) (see Fig.~\ref{fig:structure_fig}(b)), and $U_{2\pi/3}$ denotes a $2\pi/3$ rotation about the crystallographic $c$-axis perpendicular to the honeycomb plane. The most general form of $H_{\text{XXZ},\alpha}^{(1)}$ consistent with the symmetry of BCAO is given by Eq.~\eqref{eq: general coupling matrix CF}. The simpler four parameter form used here is based on Ref.~\cite{Das2021XYCobaltates}. In this model the global $U(1)$ symmetry of the pure $XXZ$ model is broken by the $D$ and $E$ terms. These are assumed to be small albeit finite to open a gap in the magnetic excitation spectrum as observed in this work (see Fig.~\ref{fig:hys_inel_fig}(a) and Fig.~\ref{fig:hys_h0l_gap}). 

It is important to note that both models can be described in either the crystallographic frame (CF) or the Kitaev frame (KF). The coordinate transformation is described in Appendix \ref{appendix: CF to KF mapping}. While the most general bi-linear nearest-neighbor spin Hamiltonian consistent with the space group symmetry involves six parameters, the $JK\Gamma\Gamma'$ and \XXZ models are different approximations with just four parameters each. As a result the $JK\Gamma\Gamma'$ interaction is most conveniently represented in the Kitaev frame (Eq.~\ref{eq:JKGGp_matrix}) while the \XXZ model is more conveniently expressed in the crystallographic frame (Eq.~\ref{eq: coupling matrices XXZ-J1-J3}). Expressing the XXZ nearest-neighbor interaction in the Kitaev frame results in $K=D-\sqrt{2}F\approx 0$ (Eq.~\ref{eq: coupling constants for XXZJ1J3 in the KF}). Thus the two models as defined represent distinct limits for the nearest-neighbor interactions. In the following, we critically examine both proposals. To discriminate between the two, we constrain the microscopic coupling constants and determine how accurately they can reproduce experimental observations such as the INS spectrum, ground state magnetic order, and magnetization.

It is also assumed that the essential physics of BCAO can be captured by a purely two-dimensional model. This assumption can be tested using LSWT. Using the dispersion from $A$-$\Gamma_1$ (Fig. \ref{fig:hys_inel_fig}, column 1), we place an upper limit on the interlayer coupling $J_1^{\prime}$. As can be seen in Fig. \ref{fig:hys_inel_fig}, there is no noticeable dispersion along this path. By tuning $J_1^{\prime}$ in LSWT, we find an upper limit of $|J_1^{\prime}|\approx0.3$~meV, at which point the dispersion predicted in LSWT exceeds the instrumental resolution of $\Delta\hbar\omega=0.06$~meV (FWHM). We therefore may reasonably model BCAO as a purely two-dimensional spin system.

\begin{figure*}
    \centering
    \includegraphics[width=1.0\textwidth]{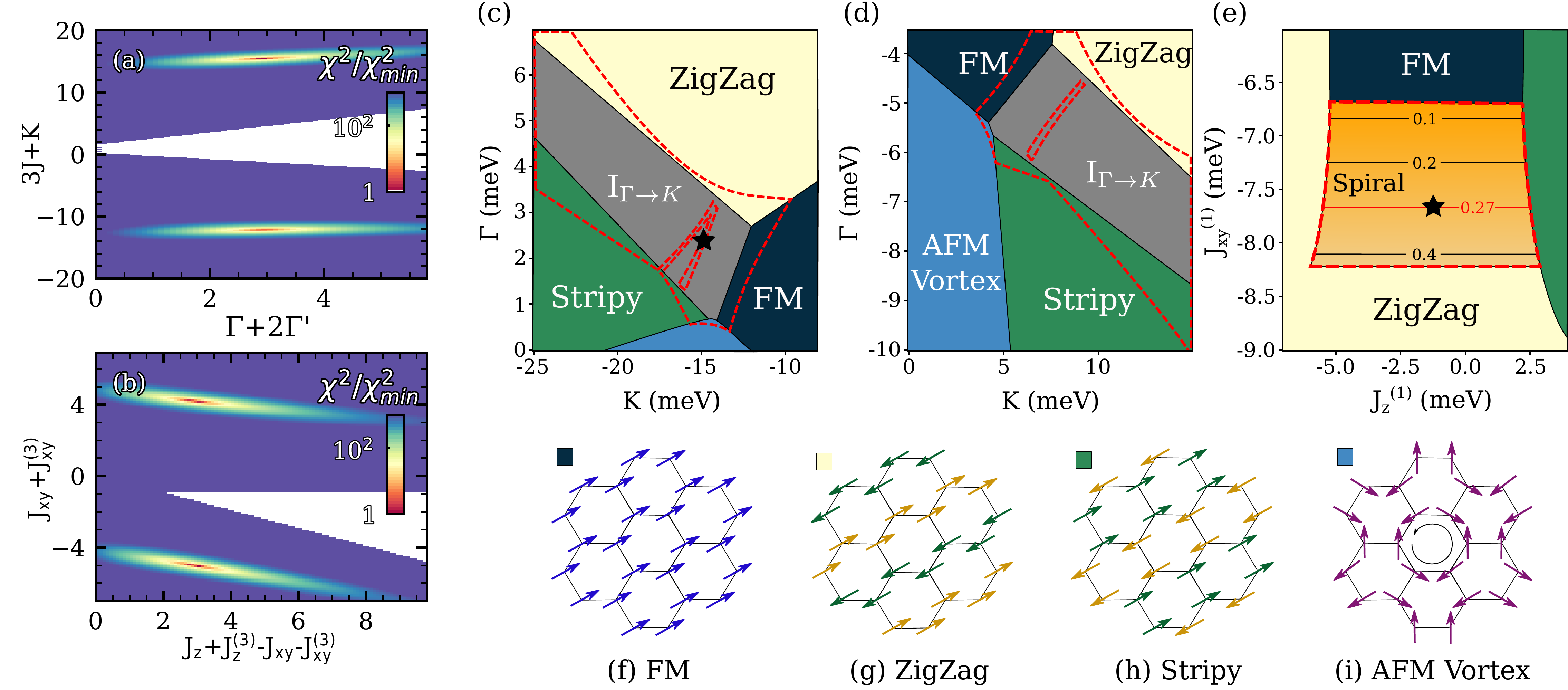}
    \caption{ (a)-(b) Normalized goodness of fit $\chi^{2}$ for the field dependence of the dispersion at the $\Gamma$ point of the \JKG and \XXZ models. (c)-(e) Classical phase diagram around the only regions where incommensurate order was observed for the restricted parameter space that reproduces the field-dependent dispersion at the $\Gamma$ point. (c) and (d) correspond to the constraints $\Gamma+2 \Gamma^{\prime}=3.0$~meV and $3 J+K=-12.1$~meV, whereas (e) corresponds to $J_{x y}^{(1)}+J_{x y}^{(3)}=-5.0$~meV and $J_{z}^{(1)}+J_{z}^{(3)}=-2.0$~meV. The red dashed lines enclose the regions where the Luttinger-Tisza approximation fails, and the ground state is instead determined by a combination of simulated annealing and a variational single-Q Ansatz. The contour lines in the spiral phase of (e) represent the magnitude of the ordering wavevector. The black stars in panels (c) and (e) are the representative points used for both models (see Eqs.~\eqref{eq: representative points JKGGp} and \eqref{eq: representative points XXZJ1J3}). (f)-(i) Representative spin configurations for the reported classical commensurate phases and the corresponding color in the phase diagrams. \label{fig:lt_phase_diagram}}
\end{figure*}

\subsection{\label{subsec: Field dependence of the gap in the polarized regime} Field dependence of the magnon dispersion in the polarized regime}

To fit a model with a large number of free parameters, it is useful to identify constraints that restrict the problem to a smaller region of phase space. In our case, a straightforward method to initially restrict the set of coupling constants to an experimentally relevant subspace is to fit the field dependence of the $\Gamma$ point magnon gap in the polarized regime illustrated in Fig.~\ref{fig:hys_inel_fig2}~(e), and the position of the second band at the zone center as seen in Fig.~\ref{fig:hys_inel_fig}~(e). Using linear spin-wave theory (LSWT), we find that the energies of the two magnon bands in the field-polarized state at the $\Gamma$ point are 
\begin{widetext}
\begin{subequations}
\begin{align}
    E_{JK\Gamma\Gamma'}^{(1)}(\mathbf{k}=\vec{0}) &= \sqrt{|\mathbf{h}|^{2}+3 S (\Gamma+2 \Gamma') |\mathbf{h}|} \\
    E_{JK\Gamma\Gamma'}^{(2)}(\mathbf{k}=\vec{0}) &= \sqrt{|\mathbf{h}|^{2}+2 S^{2}(3 J+K -\Gamma-2 \Gamma')(6 J+2 K+\Gamma+2 \Gamma')-S(12 J+4 K - \Gamma - 2 \Gamma') |\mathbf{h}|}
\end{align}
\end{subequations}
and 
\begin{subequations}
\begin{align}
    E_{\text{XXZ}\text{-}J_{1}\text{-}J_{3}}^{(1)}(\mathbf{k}=\vec{0}) &=  \sqrt{|\mathbf{h}|^2 + 3S(J_{z}^{(1)}+J_{z}^{(3)} -J_{xy}^{(1)}-J_{xy}^{(3)})|\mathbf{h}| } \\
    E_{\text{XXZ}\text{-}J_{1}\text{-}J_{3}}^{(2)}(\mathbf{k}=\vec{0}) &=  \sqrt{\left(6(J_{xy}^{(1)}+J_{xy}^{(3)}) S - |\mathbf{h}|\right) \left(3(J_{xy} + J_{xy}^{(3)}+ J_{z} + J_{z}^{(3)}) S - |\mathbf{h}|\right) }
\end{align}
\end{subequations}
\end{widetext}
for the \JKG and \XXZ models respectively. Here $\mathbf{h}=g\mu_B \mathbf{B}$ and we use the \emph{ab initio} determined $g-$factors $g_{ab}=5$ and $g_{c}=2$ \cite{Das2021XYCobaltates} for in-plane and out-of-plane fields respectively. It should be noted that the magnon dispersion at zone center for both models only depends on two specific combinations of the couplings: $3J+K$ and $\Gamma+2\Gamma'$ for the \JKG model. For the \XXZ model the gaps versus field are controlled by $J_{xy}^{(1)}+J_{xy}^{(3)}$ and $J_{z}^{(1)}+J_{z}^{(3)}$ but is independent of $D$ and $E$, which must therefore be constrained by other data.

To find a specific combination of couplings that agrees with the field-dependent position of the two bands, we introduce the goodness of fit measure
\begin{align}
    \chi^{2}=\frac{\chi^{2}_{1}+\chi^{2}_{2}}{2}
\end{align}
where $\chi^{2}_{i}$ is the weighted residual sum of squares for the $i^{th}$ band at the $\Gamma$ point for all experimental values of the magnetic field $|\vec{B}_n|$ in the polarized regime
\begin{align}
\chi^{2}_{i}= \sum_{n}\frac{\left(E_{i,\vec{B}_n}^{\text{exp}}(\vec{0})-E_{i,\vec{B}_n}^{\text{fit}}(\vec{0})\right)^{2}}{(\Delta E_{i,\vec{B}_n}^{\text{exp}})^2}, 
\end{align}
and $\Delta E_{i,\vec{B}_n}^{\text{exp}}$ is the experimental uncertainty in the position of the bands as estimated from a Lorentzian fit to the inelastic neutron scattering peaks (Fig.~\ref{fig:hys_inel_fig2}(a-d)). The resulting map of the normalized $\chi^2$ is presented in Fig.~\ref{fig:lt_phase_diagram} (a) and (b) for the two models. For both, we obtain only two localized regions that give an extremely faithful reproduction of the field dependence of the dispersion at the zone center. These regions are given by
\begin{subequations} \label{eq: constraints from field dep. JKGGp}
\begin{align} 
    &\begin{cases}  
        3J + K = -12.1(1)\text{meV}\\
        \Gamma + 2 \Gamma' = 3.0(1)~\text{meV}
    \end{cases}\\
    & \text{and} \nonumber \\
    & \begin{cases} 
        3J + K = 15.2(1)~\text{meV}\\
        \Gamma + 2 \Gamma' = 3.0(1)~\text{meV}
    \end{cases}
\end{align}
\end{subequations}
for the \JKG model, and 
\begin{subequations} \label{eq: constraints from field dep. XXZ-J1-J3}
\begin{align} 
    &\begin{cases}
        J_{xy}^{(1)} + J_{xy}^{(3)} = -5.0(1) \text{meV}  \\
    J_{z}^{(1)} + J_{z}^{(3)} = -2.0(1)~\text{meV} 
    \end{cases}\\
    &\text{and} \nonumber \\
    &\begin{cases}
        J_{xy}^{(1)} + J_{xy}^{(3)} = 4.0(1)~\text{meV}  \\
    J_{z}^{(1)} + J_{z}^{(3)} = 7.0(1)~\text{meV}.
    \end{cases}
\end{align}
\end{subequations}
for the \XXZ model. We restrict the rest of our analysis to these four sets of constraints. 

\subsection{\label{subsec: magnetic ground state}Magnetic Ground State}

To further constrain the remaining parameters, we investigate whether the two models can reproduce the experimentally observed incommensurate magnetic order within this restricted parameter set. We note that the magnetic ground state of BCAO is not an entirely experimentally settled issue. The zero-field ordered state was long-accepted to be a simple spiral structure with incommensurate ordering wavevector between the $\Gamma$ and $M$ points \cite{Regnault1977MagneticBaCo2AsO42}. However, a recent spherical neutron polarimetry study \cite{Regnault2018Polarized-neutronMagnet} suggests that the magnetic structure may be better described by double-zigzag spin-chains forming a $	\uparrow 	\uparrow 	\downarrow 	\downarrow$ pattern with some small out-of-plane canting angle. These two cases are challenging to differentiate in neutron diffraction, and both are strongly frustrated incommensurate structures. In any case, these two ground states are most likely strongly competing. In this work, we assume the incommensurate spiral order for the sake of simplicity.

To obtain a classical magnetic phase diagram, we employ the Luttinger-Tisza approximation \cite{luttinger1946theory, litvin1974luttinger}, in which a direct solution of the classical model can be obtained by relaxing the constraint of fixed spin length at every site. In regions where the Luttinger-Tisza approximation fails (i.e., the resulting solution does not respect the hard constraints on the spin length on all sites), we have employed a combination of simulated annealing on finite clusters with $2 \cdot 26^{2}$ sites, and variational single-Q Ansatz of the form
\begin{align}
    \vec{S}_{{i}}=\sqrt{1-\alpha_{i}^{2}}\left[ \cos (\vec{Q} \cdot \vec{r}_{i})\hat{\vec{e}}_{i}^{x} + \sin (\vec{Q} \cdot \vec{r}_{i}) \hat{\vec{e}}_{i}^{y} \right]+\alpha_{i} \hat{\vec{e}}_{i}^{z},
\end{align}
where the canting out of the rotation plane $\alpha_{i}$ and the orthonormal frames ($\hat{\vec{e}}_{i}^{x},\hat{\vec{e}}_{i}^{y},\hat{\vec{e}}_{i}^{z}$) are sublattice-dependent variational parameters. An extensive classical phase diagram for all reasonable values of the coupling constants that respect the constraints~\eqref{eq: constraints from field dep. JKGGp} and~\eqref{eq: constraints from field dep. XXZ-J1-J3} was performed by varying the two largest remaining free parameters for both models (i.e., $J$ and $\Gamma$ for the \JKG model, and $J_{xy}^{(1)}$ and $J_{z}^{(1)}$ for the \XXZ model). We will present this extensive phase diagram in an upcoming publication. Here, we report the experimentally relevant regions where incommensurate magnetic order was observed for the \JKG model in Fig.~\ref{fig:lt_phase_diagram}(c) and (d), and for the \XXZ model in Fig.~\ref{fig:lt_phase_diagram}(e).

For the \JKG model, we find two regions that support a ground state where the ordering wavevector is between the $\Gamma$ and $K$ points. This phase is represented in grey in Fig.~\ref{fig:lt_phase_diagram}(c) and (d) and we label it as I$_{\Gamma\to K}$. Such a magnetic order is not pertinent to the ground state of BCAO where the ordering wavevector is between the $\Gamma$ and $M$ points. To investigate if this incommensurate spiral order could be stabilized by the introduction of a third nearest-neighbor isotropic interaction $J_3$, we updated the constraints presented in Eqs. \eqref{eq: constraints from field dep. JKGGp} to account for this new coupling and obtained the corresponding phase diagram by varying $K$ and $\Gamma$ as presented in fig. \ref{fig:lt_phase_diagram} (c) and (d) for $J_3$ ranging from -5~meV to 5~meV. The details of this analysis will be presented in a future publication. For all values of $J_3$, the right $\Gamma$ to $M$ ordering is not observed in the presence of a significant Kitaev term.  Thus, we find no values of the coupling constants for the \JKG model with large Kitaev interactions that can reproduce the field dependence of the dispersion at the $\Gamma$ point and the right classical ground state even with the addition of further nearest-neighbor Heisenberg interactions. This is already a cogent indication that the \JKG model with large bond-dependent interactions may not provide an accurate description of BCAO. To provide further support to this claim, we investigate if it can correctly reproduce our inelastic neutron scattering measurements. 

In contrast, it has been reported that the \XXZ model on the honeycomb lattice supports spiral incommensurate order with an ordering wavevector that smoothly interpolates between the $\Gamma$ and $M$ points with a specific magnitude of $|\vec{k}_c|$=0.27 for $J_{xy}^{(3)}/|J_{xy}^{(1)}| \approx 0.34$ and $J_{xy}^{(1)}<0$ \cite{Fouet2001AnLattice}. This incommensurate spiral region is represented in Fig.~\ref{fig:lt_phase_diagram}(e), where there exists a line that reproduces the experimentally determined incommensurate wavevector between zigzag and stripy phases that are stabilized for large values of $|J_{z}^{(1)}|$. This spiral phase with $|\vec{k}_c|$=0.27 is also stable to the addition of anisotropic terms provided they are sufficiently small (i.e., approximately if $|D|,|E|<0.2$~meV). It is consequently natural to account for both the field dependence of the gap and the incommensurate magnetic structure of BCAO with the \XXZ model. 

\begin{figure*}
    \centering
    \includegraphics[width=\columnwidth]{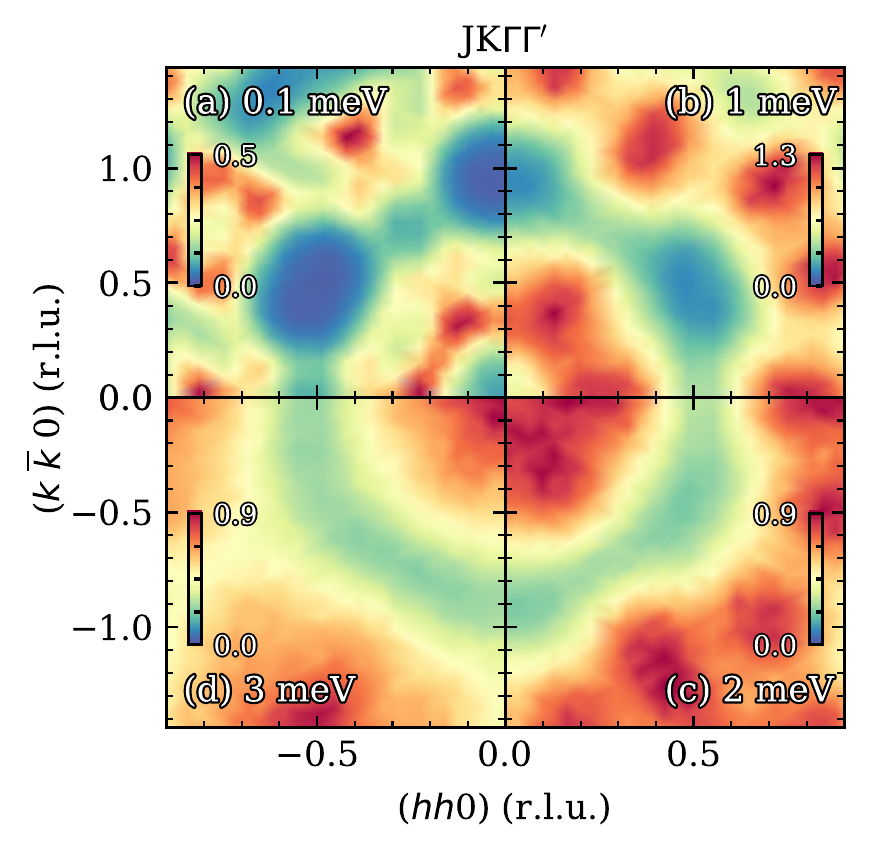}
    \includegraphics[width=1.0\columnwidth]{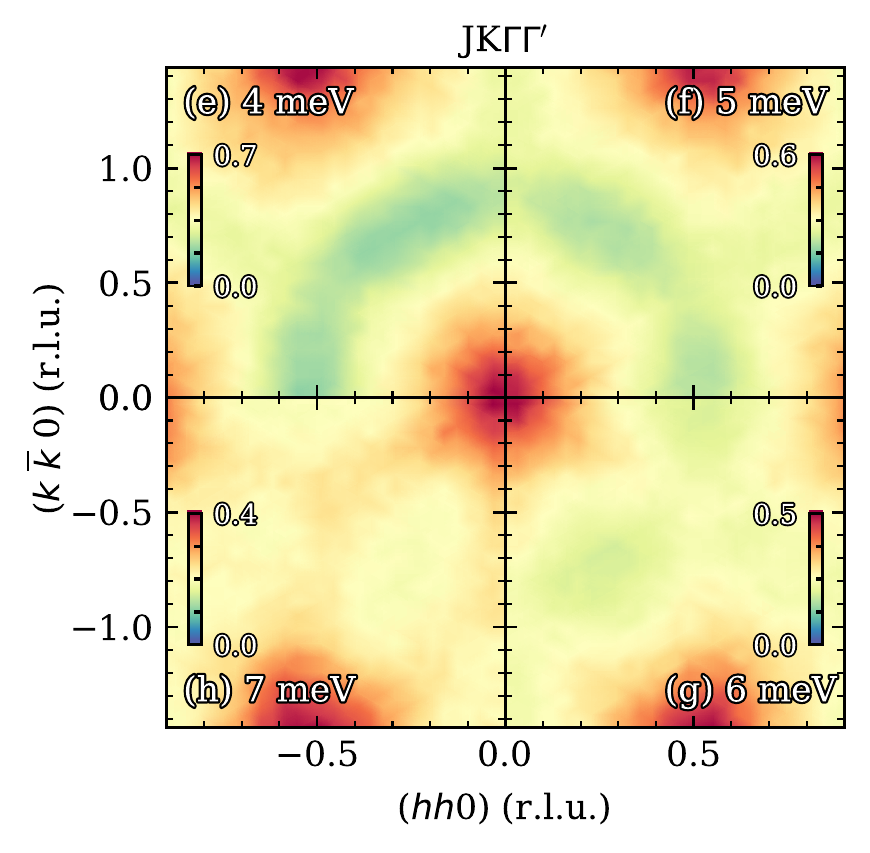}
    \includegraphics[width=1.0\columnwidth]{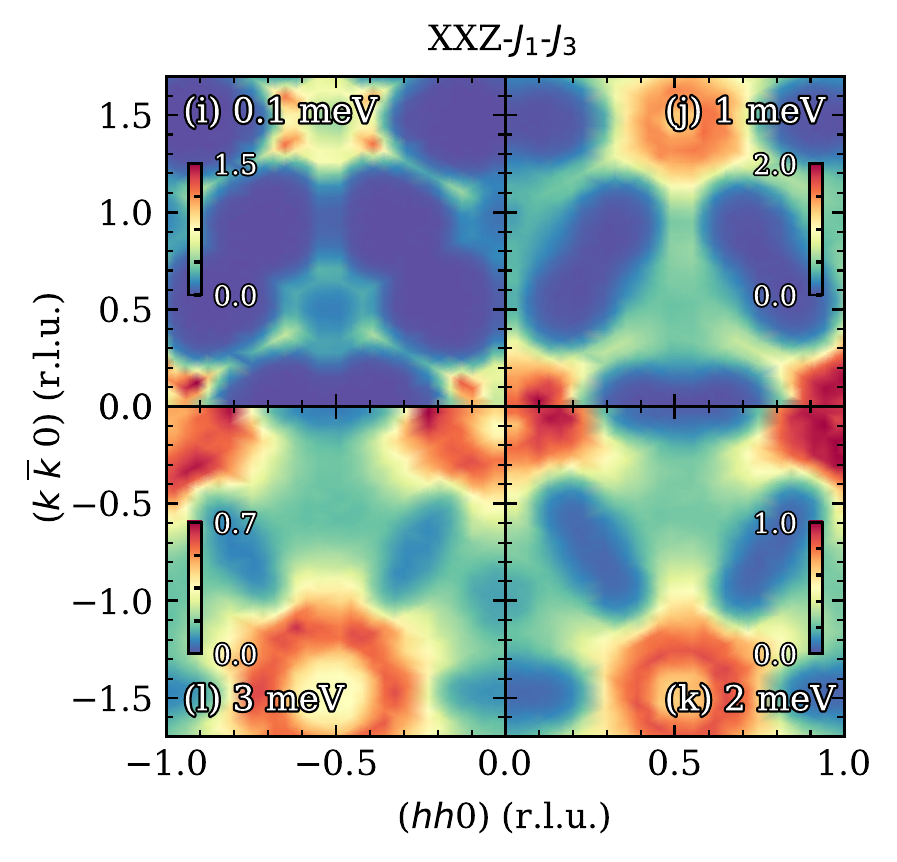}
    \includegraphics[width=1.0\columnwidth]{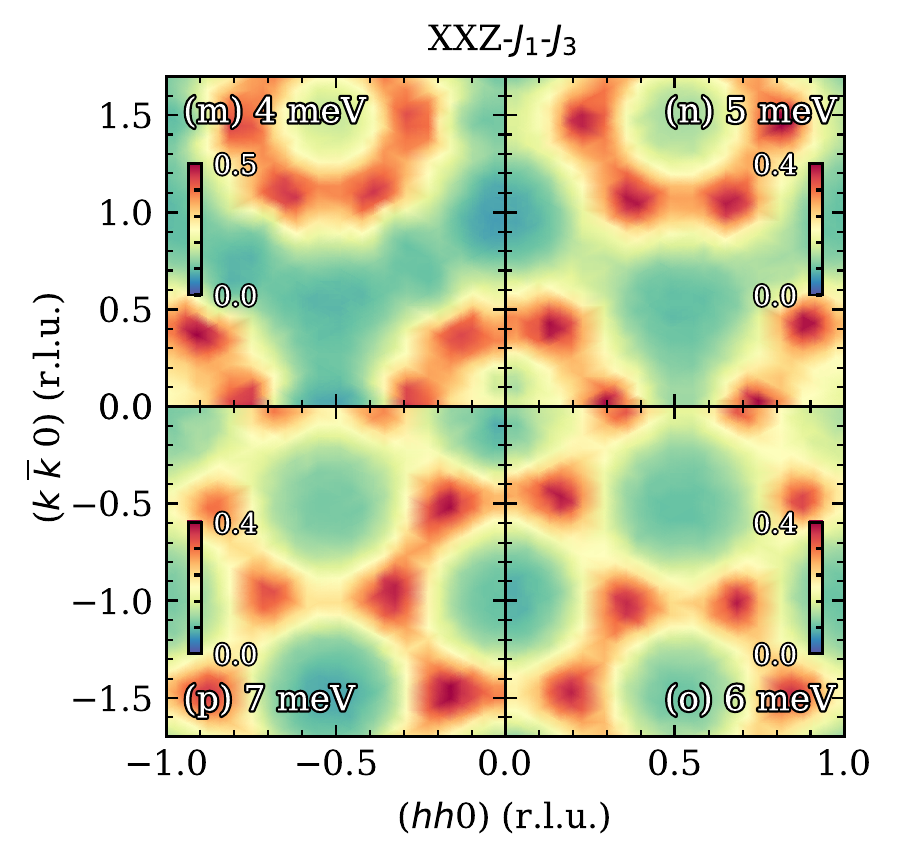}
    \caption{Constant energy slices of the dynamical spin structure factor ${\cal S}(Q,\omega)$ calculated by molecular dynamics at $T=2$ K. Color scales are arbitrary but consistent within each model. Subplots (a-h) show constant energy slices for the \JKG model, and subplots (i-p) for the \XXZ model with the parameter set presented in Eqs.~\eqref{eq: representative points JKGGp} and \eqref{eq: representative points XXZJ1J3} respectively.}
    \label{fig:dssf_md_fig}
\end{figure*}

\subsection{\label{subsec: Molecular Dynamics}Molecular Dynamics}
Within this restricted parameter space, we can compute the dynamical spin structure factors (DSSF) of both models to directly compare against the inelastic neutron scattering results. The momentum and energy-dependent dynamic structure factor is given by
\begin{align}
    \mathcal{S}^{\mu\nu}(\mathbf{q}, \omega)=\frac{1}{2\pi N}\sum_{i,j}^{N}\int \mathrm{d}t\  e^{-i\mathbf{q}\cdot(\mathbf{r}_i-\mathbf{r}_j)+i\omega t}\langle S^\mu_i(t)S^\nu_j(0)\rangle
\end{align}
where $N$ is the number of lattice sites. We investigated the spectrum with unpolarized neutrons, described by 
\begin{align}
    \mathcal{S}(\mathbf{q}, \omega)=\sum_{\mu,\nu}\left[\hat{z}_\mu\cdot\hat{z}_\nu-\frac{(\hat{z}_\mu\cdot \mathbf{q})(\hat{z}_\nu\cdot\mathbf{q})}{q^2}\right]\mathcal{S}^{\mu\nu}(\mathbf{q}, \omega).
\end{align}
Here, $\hat{z}_\mu$ are the basis vectors for the KF for the \JKG model or the CF for the \XXZ model. We study the classical limits of both models using finite temperature Monte Carlo and obtain the DSSF using molecular dynamics (MD). The details of the numerical techniques can be found in Appendix \ref{appendix: Molecular Dynamics}. 

Based on constraints described in Sections \ref{subsec: Field dependence of the gap in the polarized regime} and \ref{subsec: magnetic ground state}, we choose the parameters
\begin{gather}
\label{eq: representative points XXZJ1J3}
\begin{aligned}
J_{x y}^{(1)}&=-7.6~\mathrm{meV}, \\
J_{z}^{(1)}&=-1.2~\mathrm{meV},\\
J_{x y}^{(3)}&=2.5~\mathrm{meV}, \\
J_{z}^{(3)}&=-0.85~\mathrm{meV}, \\
D&=0.1~\mathrm{meV},\\
E&=-0.1~\mathrm{meV}
\end{aligned}
\end{gather}
and
\begin{gather} \label{eq: representative points JKGGp}
\begin{aligned}
J&=0.97~\mathrm{meV}, \\
K&=-15.0~\mathrm{meV},\\
\Gamma&=2.5~\mathrm{meV},  \\
\Gamma^{\prime}&=0.25~\mathrm{meV}
\end{aligned}    
\end{gather}
as the best-fit for both models. We note that Eq.~\eqref{eq: representative points XXZJ1J3} may equivalently be written in the KF and that Eq.~\eqref{eq: representative points JKGGp} may be written in the CF. The corresponding Hamiltonian parameters are given in Appendix \ref{appendix: CF to KF mapping}. The two models support the incommensurate spiral and I$_{\Gamma\to K}$ phases respectively, as depicted in \fig{fig:lt_phase_diagram}. For \XXZ, The remaining free parameters $J_{z}^{(1)}$, $D$ and $E$ were chosen to be small enough in magnitude to yield the correct spiral order, approximately reproduce the gap observed at zero field in Fig.~\ref{fig:hys_inel_fig}(a), and provide the best reproduction of the neutron scattering in the polarized regime as presented in Fig.~\ref{fig:hys_inel_fig}(c-e). However, even with our tight constraints on the allowed parameter sets, we note that there may be other proximate parameter sets that may yield similar results. The parameters for the \JKG model were also chosen to provide the best fit to the neutron scattering data in the polarized regime within the I$_{\Gamma\to K}$ phase.
\begin{figure}[t!]
    \centering
    \includegraphics[width=\columnwidth]{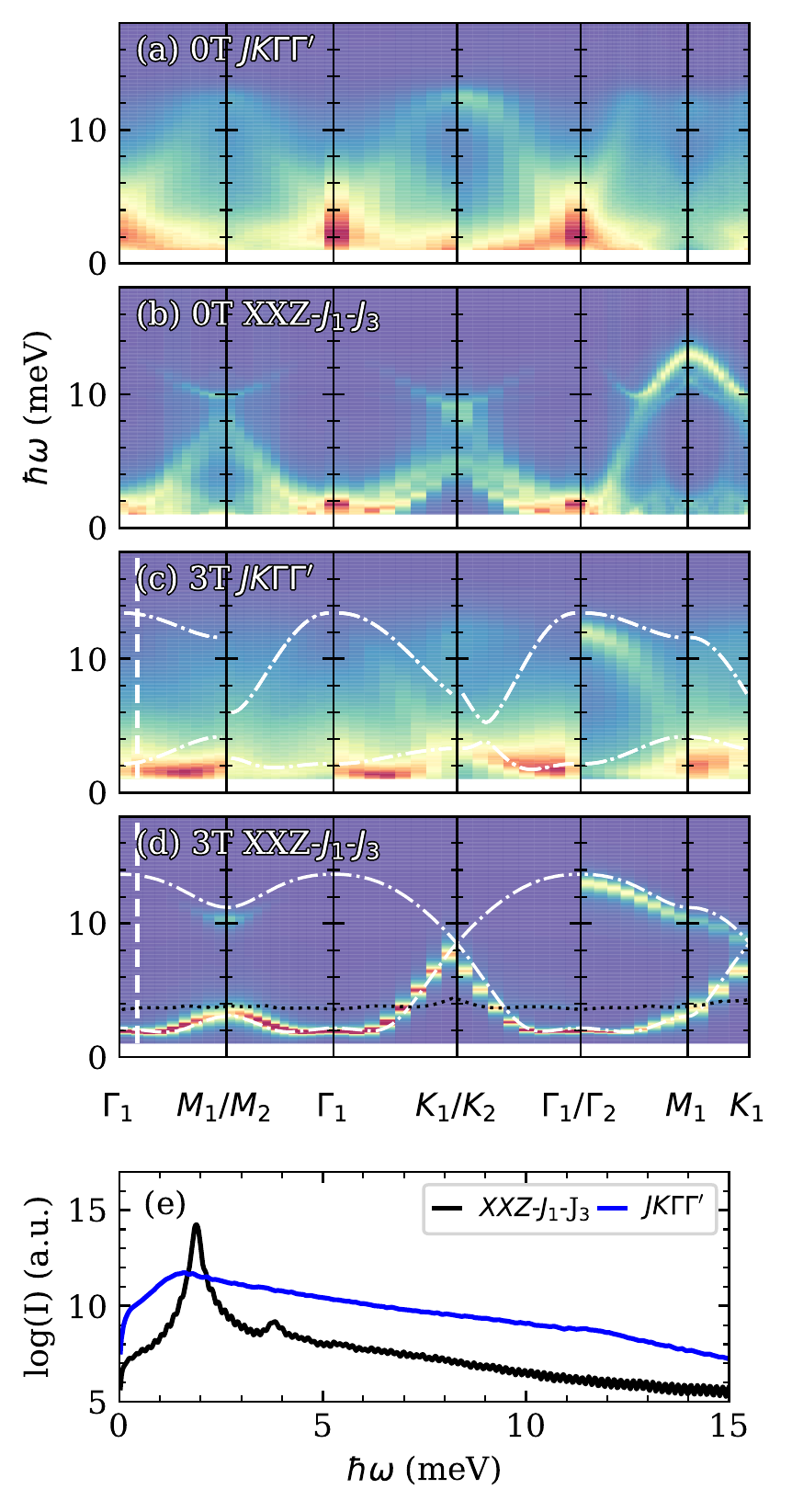}
    \caption{Dynamical spin-structure factor obtained by MD between high symmetry points for both the \JKG model (a,c) and the \XXZ model (b,d) with an in-plane field of 0~T at $T=1$~K and 3~T at $T=2$~K, respectively. Parameters used for both models are given by Eq.~\eqref{eq: representative points JKGGp} and Eq.~\eqref{eq: representative points XXZJ1J3}. The LSWT dispersion for each model and set of parameters are overlaid in (c) and (d) using the dashed dotted line. For the \XXZ case the dotted black line represents the lower edge of the two-magnon continuum calculated using the LSWT dispersion. (e) Intensity cut near the $\Gamma$ point (along the vertical dashed line in (c) and (d)) for both models at 3~T.}
    \label{fig:dssf_h00_md}
\end{figure}

Fig.~\ref{fig:dssf_md_fig} shows the constant energy slices of the DSSF for the \JKG (Fig.~\ref{fig:dssf_md_fig}(a-h)) and the \XXZ (Fig.~\ref{fig:dssf_md_fig}(i-p)) models, which may be compared to the experimental results in Fig.~\ref{fig:macs15K_inel_fig}. The \JKG model does not reproduce any qualitative features seen in the in-plane scattering experiments apart from a buildup of spectral weight at the $\Gamma$ point due to the ferromagnetic nature of the dominant Kitaev term. In contrast, the \XXZ model is remarkably able to reproduce many features in the higher energy cuts. The resemblance is especially striking when comparing the DSSF around 2~meV and above with the 15~K experimental measurements where the observed six-fold rotationally symmetric high-intensity hexagonal structure joined by low-intensity oval pockets is clearly reproduced in the DSSF. It should be emphasized that even though the underlying spiral magnetic order is consistent with BCAO, it is highly non-trivial that the excitation spectrum also agrees, deeming these results even more noteworthy. We also note that the better agreement between numerical results and high-temperature data is expected since we are dealing with purely classical two-dimensional simulations. As an alternative point of view, we also present the DSSF along a path in the first Brillouin zone for both models in Fig.~\ref{fig:dssf_h00_md}(a) and (b).

Next, we present the DSSF as a function of energy between high symmetry points in the field-polarized regime at $\mu_{0}H=3$~T and at zero-field with the LSWT dispersion overlaid in \fig{fig:dssf_h00_md} (c)-(d). The MD results for the \XXZ model with our parameters show good agreement with experimental measurements reported in Fig.~\ref{fig:hys_inel_fig}. As can be seen from the intensity cut near the zone center in Fig.~\ref{fig:dssf_h00_md}(e), MD simulations even capture the two-magnon mode reported in Fig.~\ref{fig:hys_inel_fig2}(d) and THz spectroscopy \cite{Zhang2021In-BaCo_2AsO_4_2,Shi2021MagneticSpectroscopy}. In contrast, the experimental results do not seem to be well described by our parameter set for the \JKG model. LSWT for this model predicts a magnon dispersion that differs significantly between the $\Gamma$ and $M_1$ points compared to the $\Gamma$ and M$_2$ points. Such a large anisotropy is incompatible with our INS measurements presented in Fig.~\ref{fig:hys_inel_fig}. MD further predicts extremely blurry bands as highlighted by the intensity cut near the $\Gamma$ point presented in Fig.~\ref{fig:dssf_h00_md}(e). This blurriness of the bands in the \JKG model reflects the short magnon lifetime due to the large frustration of the system. As such, the sharp magnon bands observed experimentally are much more compatible with the \XXZ model. Since LSWT neglects all magnon-magnon interactions, such information would have been completely missed by only employing this approach, emphasizing the need to go beyond LSWT when investigating highly frustrated magnets beyond the field-polarized regime. 

The magnon dispersions predicted from LSWT theory and MD for the \XXZ model seem to be in close correspondence for the whole scattering path considered. The most meaningful discrepancy is around the $\Gamma$ point where the LSWT magnon band is concave down with a minimum around the incommensurate ordering wavevector, whereas the MD bands are much more flat and concave up as the experimental data for BCAO (Fig.~\ref{fig:hys_inel_fig}). This disagreement between LSWT and MD indicates the presence of important nonlinearities, which are taken into account in MD but ignored in LSWT, and which lead to a significant renormalization of the magnon dispersion. Indeed we observe that for larger values of the magnetic field, the negative concavity of the LSWT is progressively suppressed until it is concave up as observed in MD (see supplementary information). For such large fields, the relative importance of nonlinearities is progressively suppressed and thus MD and LSWT are expected to be in excellent agreement. This comparison between predictions from MD and LSWT suggest that the disagreement for the concavity of the dispersion at the zone center observed experimentally and predicted from LSWT can be accounted for by the presence of strong nonlinear effects.

\subsection{\label{subsec: In-plane Magnetization} In-Plane Magnetization}
\begin{figure}[h!]
    \centering
    \includegraphics[width=1.0\columnwidth]{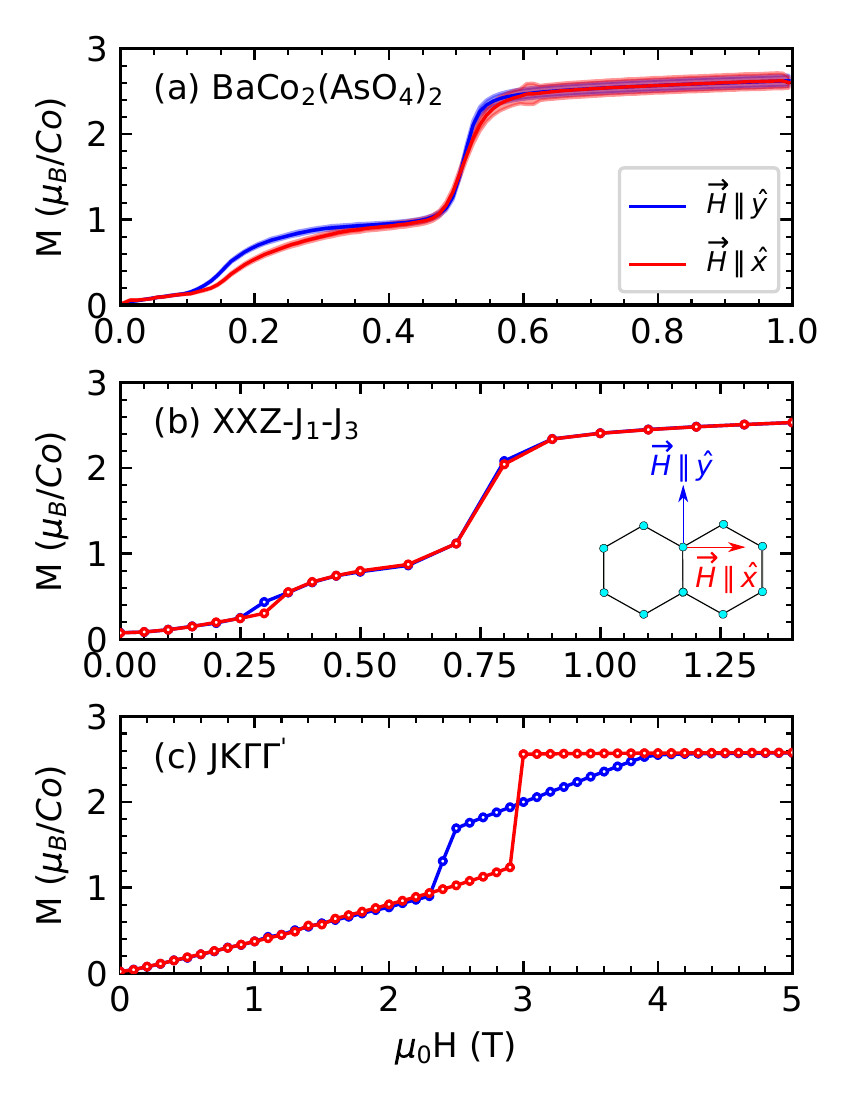}
    \caption{(a) In-plane angle-dependent magnetization for $\rm BaCo_2(AsO_4)_2$ measured on a small single crystal sample. The red (blue) curve corresponds to a field orientation along the $x$-axis ($y$-axis) as depicted in the inset. For clarity, the magnetization curves are only shown for an increasing field from 0 T. (b) Magnetization curves calculated using Monte Carlo methods for the \XXZ model with the set of exchange parameters given in Eq.~\eqref{eq: representative points XXZJ1J3} at a temperature of $T=0.695$ K, and (c) for the \JKG model with the parameter set in Eq.~\eqref{eq: representative points JKGGp} ($T=0$~K). 
    } 
    \label{fig:md_magnetization}
\end{figure}

As a final comparison between the two models and experiments, we examine the in-plane magnetization for different magnetic field orientations. Increasing the field from zero, three successive magnetization plateaus can be observed in Fig.~\ref{fig:md_magnetization}(a), where the transitions from one plateau to the next correspond to the field-induced magnetic phases transitions at $H_{c1}$ and $H_{c2}$ respectively. Measuring the magnetization with magnetic fields along the $x$ and $y$-axis of the CF (see Fig.~\ref{fig:structure_fig}(b)), only a very small anisotropy is observed around the first and second transitions. Using our parameter set for the \XXZ model, the qualitative resemblance between the magnetization curves predicted from finite temperature Monte Carlo at $T=0.695$~K presented in Fig.~\ref{fig:md_magnetization}(b) and measurements is quite striking. The three successive plateaus are clearly observed, and the small anisotropy between the curves for two different field orientations around the two transitions is reproduced. We have verified that the intermediate field-driven phase also has an ordering wavevector of magnitude $|\vec{k}_{c}|=1/3$ as observed experimentally. 

In stark contrast, as presented in Fig.~\ref{fig:md_magnetization}(c), the magnetization curves predicted from Monte Carlo at $T=0$~K for the \JKG model with our parameter set yields magnetization curves that differ significantly for the two directions. For a field along the $y$-axis, Monte Carlo for this model yields an intermediate field-induced phase transition before the transition to the field polarized phase, whereas for a field along the $x$-axis, there is a single low field transition directly to a fully polarized state. It should additionally be noted that, especially for the magnetic field along the $\hat{y}$-direction, the system transitions to the polarized regime at a much larger value of the magnetic field than in experiments. The temperatures in each case are chosen such that they most closely reproduce the experimentally observed $M(H)$ curves, the details of which are in the supplemental text.

\section{\label{sec: Discussion} Discussion and Conclusions}

Though our initial interest in BCAO was as a candidate Kitaev material, our results strongly favor the \XXZ over the Kitaev model. Firstly, we are able to differentiate between the two models by examining the zero-field ground state of BCAO using constraints from LSWT. We find that the \JKG model cannot simultaneously reproduce the field dependence of $\Gamma$ point gap and the magnetically ordered state observed in BCAO. The incommensurate spiral order reported for BCAO is also incompatible with a large ferromagnetic Kitaev interaction, even with the addition of a third nearest-neighbor Heisenberg interaction. In contrast, the spiral magnetic order is reproduced in the \XXZ picture. Perhaps the most striking contrast between the two models is seen in our calculated DSSF results presented in Fig.~\ref{fig:dssf_md_fig}. Qualitatively, it is immediately obvious that the \JKG model fails to describe the zero-field INS, whereas the \XXZ model is able to capture some highly non-trivial features. Some of these features are the result of a strong renormalization of the excitation spectra from nonlinear magnon-magnon interactions as revealed through a comparison of molecular dynamics results to LSWT. 

Lastly, the in-plane magnetization curves predicted by the \XXZ model at finite temperature are in remarkable qualitative agreement with measurements and can even reproduce subtle features such as small anisotropy observed between different field directions within the basal plane. The \JKG model yields highly anisotropic curves that cannot be reconciled with experiments, which further supports the \XXZ picture. In conclusion, our work provides compelling evidence for the description of BCAO in terms of an \XXZ model with only small bond dependant interactions. This description is able to reproduce many non-trivial measurements, where the \JKG model fails. Furthermore, we provide strong constraints on possible microscopic couplings for this model, making BCAO amenable to meaningful and specific future theoretical predictions.

Our work stresses the necessity of a critical reexamination of the proposal for Kitaev physics in cobaltates. Indeed, our results indicate that BCAO, and likely many other cobaltates, do not fall within the regime of interest that was considered in the theoretical proposals of Refs. \cite{Liu2018PseudospinModel, Sano2018Kitaev-HeisenbergInsulators} to generate large Kitaev interactions in $3d$ transition metal compounds, where direct hopping was assumed to be much weaker than ligand-assisted hopping. The strong easy-plane anisotropy we observe also highlights that local distortions of the crystal field environment play an essential role in these compounds and need to be better understood. We therefore emphasize the need for thorough \emph{ab initio} examinations of the exchange interactions of other cobaltates that were proposed as KSL candidates, and underscore the great caution required to interpret future results before making any claim about the apparent realization of Kitaev physics. As an example, by simply fitting the INS spectrum of BCAO along the $(h00)$ direction in the polarized regime with LSWT, one could have wrongly reached the conclusion that the low energy physics could be modelled by a large ferromagnetic Kitaev interaction with subdominant isotropic and off-diagonal interaction as in Eq.~\eqref{eq: representative points JKGGp}. It is imperative to ensure that the proposed model is compatible with the observed ground state order and thermodynamics measurements. Examples of potential materials that fit this call for the reassessment of magnetic interactions in honeycomb cobaltates are Na$_2$Co$_2$TeO$_6$, Na$_3$Co$_2$SbO$_6$ \cite{Yao2020FerrimagnetismNa2Co2TeO6,li2022sign,kim2021antiferromagnetic,songvilay2020kitaev}, and Li$_3$Co$_2$SbO$_6$ \cite{Vivanco2020CompetingMagnet,lin2021field}, which feature strong easy-plane anisotropy and curious field-driven transitions \cite{Yao2020FerrimagnetismNa2Co2TeO6} similar to BCAO.

As experimentally relevant methods for this critical reexamination of cobaltates, we suggest the investigation of angle-dependent magnetization and specific heat. Indeed, the \JKG model with large Kitaev interaction generically predicts significant anisotropy for thermodynamic quantities as a function of an in-plane field angle (e.g., see Fig.~\ref{fig:md_magnetization}(c)) in contrast to the \XXZ model where the in-plane magnetization is virtually independent of the field direction within the honeycomb plane \cite{gordon2021testing,janssen2019heisenberg, sears2020ferromagnetic, gohlke2018quantum}. As a point of reference, a strong in-plane anisotropy has been observed in the leading Kitaev candidate $\alpha$-RuCl$_3$  \cite{Bachus2021Angle-dependent-RuCl3,Tanaka2022ThermodynamicLiquid,modic2021scale} as well as in $\beta-$Li$_2$IrO$_3$ \cite{Majumder2019AnisotropicStudy}. These relatively simple checks are of great practical interest since they are much easier to perform than single-crystal INS in a magnetic field. For candidate cobaltates whose magnetization cannot be explained by the \XXZ or other simple models, single-crystal INS examining the polarized states of these materials may be necessary to unambiguously reveal the nature of their magnetic interactions. 

Although BCAO does not appear to realize large Kitaev-type interactions but rather an \XXZ model with competing interactions, its physical properties are nonetheless still of great interest. BCAO now presents the extremely rare case of an almost perfect two-dimensional honeycomb magnet with well-known values for its dominant exchange interactions (Eq.~\ref{eq: representative points XXZJ1J3}). The geometric frustration due to competition between first and third nearest-neighbors could imply that BCAO is in close proximity to a QSL. Previous theoretical investigations have demonstrated the possibility of purely isotropic models on the honeycomb lattice to host QSLs ground states \cite{merino2018role, zhu2014quantum, bishop2012frustrated}. Although all of these studies have reported the necessity of second nearest-neighbor interaction (which seems to be negligible for BCAO) to stabilize a QSL, it might still be conceivable to drive an XXZ system near a QSL by applying an external magnetic field. Such a possibility has, to our knowledge, not been explored yet. The spectral weight collapse with the application of an out-of-plane field reported by THz spectroscopy \cite{Zhang2021In-BaCo_2AsO_4_2} may potentially signal the onset of such a topological phase. Possibly related to this, the strong interaction between the one and two-magnon modes reported here (Fig.~\ref{fig:hys_el_fig}) is also unusual and interesting. These are promising directions for future studies of BCAO and a number of other existing honeycomb cobaltates.

\emph{Note added:} While writing this paper, we became aware of three \emph{ab initio} theoretical investigations of BCAO. Two of them \cite{maksimov2022ab, winter2022magnetic} corroborate our conclusion regarding the nature of magnetic interaction in BCAO and propose a coherent microscopic theory for it. Another \cite{Samantha2022} observes the formation of spin-orbit-entangled $J_{\text{eff}}$=1/2 moments that could potentially support the claim of Kitaev physics. Ref. \cite{maksimov2022ab} importantly finds that the double-zigzag state is energetically competitive with the incommensurate spiral state (classical ground state as shown in our work), and can be stabilized by quantum fluctuations or relaxation of local atomic positions. 

\begin{acknowledgments}
We gratefully acknowledge valuable discussions with Peter Armitage, Hae-Young Kee, Xinshu Zhang, and Sreekar Voleti. This work was supported as part of the Institute for Quantum Matter, an Energy Frontier Research Center funded by the U.S. Department of Energy, Office of Science, Basic Energy Sciences under Award No. DE-SC0019331. C.B. was supported by the Gordon and Betty Moore foundation EPIQS program under GBMF9456. F.D., E.Z.Z., and Y.B.K. were supported by the NSERC of Canada and the Center for Quantum Materials at the University of Toronto. E.Z.Z. was further supported by the NSERC Canada Graduate Scholarships-Doctoral (CGS-D). The research at the ORNL Spallation Neutron Source was sponsored by the U.S. Department of Energy, Office of Basic Energy Sciences. A portion of this research used resources at the Spallation Neutron Source, a DOE Office of Science User Facility operated by the Oak Ridge National Laboratory. Most of the computations were performed on the Cedar and Niagara clusters, which are hosted by WestGrid and SciNet in partnership with Compute Canada.
\end{acknowledgments}

\bibliography{bcao_refs_2}

\appendix

\section{\label{appendix: magnetic structure}Magnetic Structure}

\begin{figure}[b]
    \centering
    \includegraphics[width=1.0\columnwidth]{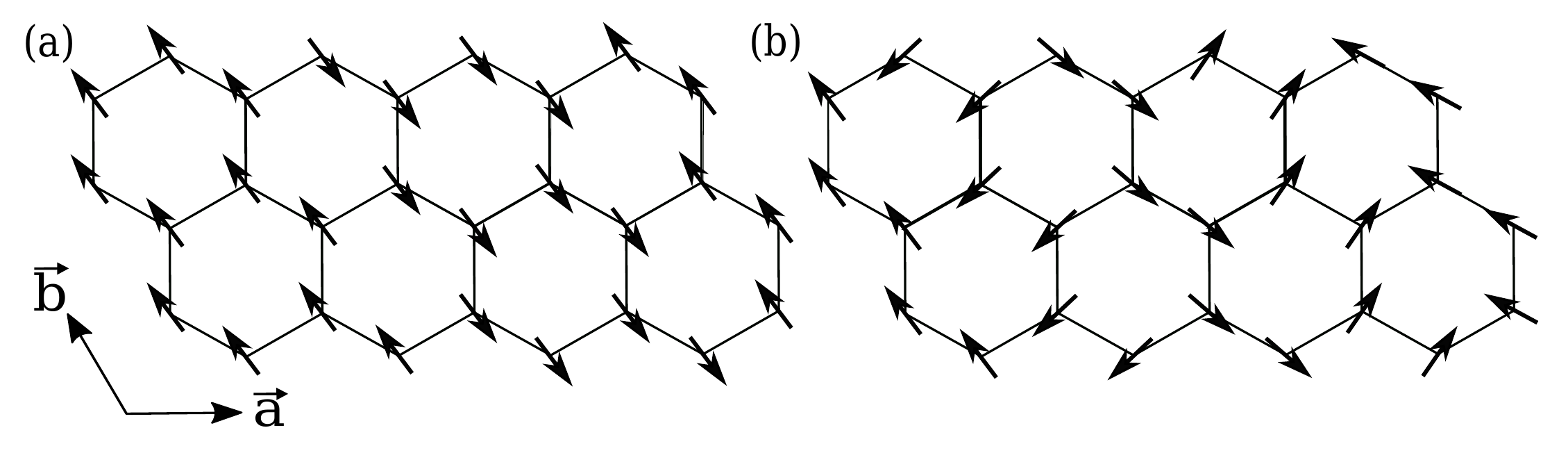}
    \caption{Real space sketches of the two reported spin structures in BCAO. (a) Shows the double zigzag structure reported in Ref. \cite{Regnault2018Polarized-neutronMagnet}, whereas (b) schematically presents the previously reported spiral structure. We do not differentiate between sublattices for the purpose of this sketch. }
    \label{fig:mag_struct_sketch}
\end{figure}

Detailed studies of the magnetic structure of BCAO have been performed using polarized neutron scattering \cite{Regnault2018Polarized-neutronMagnet}, which are consistent with the present measurements. We begin with the most general model for the spiral order with an allowed canting angle out-of-plane. Assuming this type of order, the moment on each site in the spiral ordered phase may be written in the crystallographic frame as
\begin{eqnarray}
    \overrightarrow{m}_j = |S|\{\sin(\Phi)[\cos(-\textbf{k}\cdot\textbf{r}_i)]\hat{x}-\nonumber\\
    \sin(\Phi)[\sin(-\textbf{k}\cdot\textbf{r}_i)]\hat{y}+\nonumber\\
    \cos(\Phi)[\cos(-\textbf{k}\cdot\textbf{r}_i)]\hat{z}\}.
    \label{eq:moment_simple}
\end{eqnarray}
While the spiral state was the long-accepted picture of the low-temperature magnetic order in BCAO, recent spherical neutron polarimetry studies \cite{Regnault2018Polarized-neutronMagnet} find that the magnetic structure is better described as a collinear square wave structure that necessarily has higher order harmonic components. 

These two cases are  difficult to differentiate in neutron diffraction. Both are strongly frustrated incommensurate structures that should be close in energy. For our theoretical analysis, we employed the simpler single-k spiral order. Sketches of both magnetic structures are given in Fig.~\ref{fig:mag_struct_sketch}(a-b).
\section{\label{appendix: CF to KF mapping} Mapping between the crystallographic and Kitaev frames}

The \XXZ model can be related to the \JKG model by rotating the CF to the KF. Written in the KF, the local basis vectors of the CF are given by 
\begin{subequations} \label{eq: basis vectors for KF}
  \begin{align}
    \hat{x} &= \frac{1}{\sqrt{6}} \left(1,1,-2\right)^{T} \\
    \hat{y}&= \frac{1}{\sqrt{2}} \left(-1 , 1 , 0 \right)^{T} \\
    \hat{z} &=  \frac{1}{\sqrt{3}} \left( 1,1,1 \right)^{T}
  \end{align}
\end{subequations}
It follows that the spins written in the CF and the KF are related by the following transformation
\begin{align}
    \mathcal{U} &= 
    \mqty(
        \frac{1}{\sqrt{6}} & -\frac{1}{\sqrt{2}} & \frac{1}{\sqrt{3}} \\
        \frac{1}{\sqrt{6}} & \frac{1}{\sqrt{2}} & \frac{1}{\sqrt{3}} \\
        -\sqrt{\frac{2}{3}} & 0 & \frac{1}{\sqrt{3}}
    ).
\end{align}
The local exchange matrices are accordingly  mapped from the CF to the KF by 
\begin{align}
    H^{(i)}_{\text{KF},\gamma} &= \mathcal{U} H^{(i)}_{\text{CF},\gamma} \mathcal{U}^{T}.
\end{align}
Mapping general bilinear couplings on the nearest-neighbor $z$ bond in the CF of the form
\begin{align} \label{eq: general coupling matrix CF}
H_{\text{CF},z}^{(1)} &=  
    \begin{pmatrix} 
        J_{xy}^{(1)} + D & E & F \\
        E & J_{xy}^{(1)} - D & G\\
        F & G & J_{z}^{(1)}
    \end{pmatrix}
\end{align}
to the KF, we obtain the exchange matrix 
\begin{align} \label{eq: general coupling matrix LKF}
H_{\text{KF},z}^{(1)}=\mqty(
J+\eta & \Gamma & \Gamma_{1}^{\prime} \\
\Gamma & J-\eta & \Gamma_{2}^{\prime} \\
\Gamma_{1}^{\prime} & \Gamma_{2}^{\prime} & J+K )
\end{align}
with the following identification
\begin{subequations} \label{eq: coupling constants for XXZJ1J3 in the KF}
  \begin{align}
    J & =-\frac{D}{3}+\frac{\sqrt{2} F}{3}+\frac{2 J_{xy}^{(1)}}{3}+\frac{J_{z}^{(1)}}{3}\\ 
    \eta & = -\frac{E}{\sqrt{3}}-\sqrt{\frac{2}{3}} G \\ 
    \Gamma & = \frac{2 D}{3}+\frac{\sqrt{2} F}{3}-\frac{J_{xy}^{(1)}}{3}+\frac{J_{z}^{(1)}}{3}\\ 
    \Gamma_{1}^{\prime} & = -\frac{D}{3}+\frac{E}{\sqrt{3}}-\frac{F}{3 \sqrt{2}}-\frac{G}{\sqrt{6}}-\frac{J_{xy}^{(1)}}{3}+\frac{J_{z}^{(1)}}{3}\\ 
    \Gamma_{2}^{\prime} & = -\frac{D}{3}-\frac{E}{\sqrt{3}}-\frac{F}{3 \sqrt{2}}+\frac{G}{\sqrt{6}}-\frac{J_{xy}^{(1)}}{3}+\frac{J_{z}^{(1)}}{3} \\ 
    K & =D - \sqrt{2} F.
    \end{align}
\end{subequations}
\newline
We may write an equivalent conversion from the KF to the CF 
\begin{subequations} \label{eq: coupling constants for XXZJ1J3 in the KF inverse}
  \begin{align}
    J_{xy}^{(1)} & = \frac{1}{3}(-\Gamma-\Gamma_{1}^{\prime}-\Gamma_{2}^{\prime}+3J+K)\\ 
    J_{z}^{(1)} & = \frac{1}{3}(2\Gamma + 2\Gamma_{1}^{\prime} +2\Gamma_{2}^{\prime} +3J+K)\\ 
    D & = \frac{1}{3}(2\Gamma-\Gamma_{1}^{\prime} -\Gamma_{2}^{\prime} +K)\\ 
    E & = \frac{\sqrt{3}}{8}(3\Gamma_{1}^{\prime} -3\Gamma_{2}^{\prime} -2\eta) \\
    F & = \frac{\sqrt{2}}{6}(2\Gamma-\Gamma_{1}^{\prime} - \Gamma_{2}^{\prime} -2K)\\
    G & = \frac{\sqrt{6}}{8}(-\Gamma_{1}^{\prime} +\Gamma_{2}^{\prime} -2\eta).
    \end{align}
\end{subequations}
It should be noted that assuming ideal edge-sharing bonds with $C_{2v}$ symmetry, the NN couplings in the KF of Eq. \eqref{eq: general coupling matrix LKF} are constrained to $\eta=0$ and $\Gamma_{1}^{\prime}=\Gamma_{2}^{\prime}$. These constraints translate to $E=G=0$ for the couplings in the CF of Eq. \eqref{eq: general coupling matrix CF}. The $R\bar{3}$ spacegroup associated with BCAO does break the $C_{2v}$ symmetry for the nearest-neighbor bond. Co occupies the 6c Wyckoff site at $(00z)$ where $z=0.17014$\cite{Zhong2020Weak-fieldHoneycomb}. This corresponds to a puckering of Co in and out of the honecomb plane by a distance $\pm(z-\frac{1}{6})c$=0.082\AA. 

The values in Eq.~\eqref{eq: representative points XXZJ1J3} in the CF are then represented in the KF by the following 
\begin{subequations} \label{eq: representative points XXZJ1J3_LKF}
\begin{align}
J^{(1)}&=-5.5~\mathrm{meV}, \\
K^{(1)}&=0.1~\mathrm{meV},\\
\eta^{(1)}&=0.06~\mathrm{meV},\\
\Gamma^{(1)}&=2.2~\mathrm{meV},  \\
\Gamma^{\prime(1)}_1&=2.0~\mathrm{meV} \\
\Gamma^{\prime(1)}_2&=2.2~\mathrm{meV} \\
J^{(3)}&=1.38~\mathrm{meV}, \\
K^{(3)}&=0.0~\mathrm{meV},\\
\eta^{(3)}&=0.0~\mathrm{meV},\\
\Gamma^{(3)}&=-1.2~\mathrm{meV},  \\
\Gamma_1^{\prime(3)}&=-1.2~\mathrm{meV}, \\
\Gamma_2^{\prime(3)}&=-1.2~\mathrm{meV}. 
\end{align}    
\end{subequations}
Here, the superscripts denote the NN and third NN bonds. Equivalently, the set of test parameters in Eq.~\eqref{eq: representative points JKGGp} in the KF may be written in the CF as
\begin{subequations} \label{eq: representative points JKGGp_GCF}
\begin{align}
J^{(1)}_{xy}&=-5.0~\mathrm{meV}, \\
J^{(1)}_{z}&=-2.0~\mathrm{meV}, \\
D&=-3.5~\mathrm{meV}, \\
E&=0~\mathrm{meV}, \\
F&=8.1~\mathrm{meV}, \\
G&=0~\mathrm{meV}. 
\end{align}    
\end{subequations}
\newline
\section{\label{appendix: Molecular Dynamics} Molecular Dynamics}

We first use finite temperature Monte Carlo (MC) techniques to obtain the spin configurations needed to compute the spin correlations. We treat the spins classically, i.e. we treat the spins as vectors $\mathbf{S}=\left(S_{x}, S_{y}, S_{z}\right)$ and we fix the magnitude to be $S=1 / 2$. We use parallel tempering to first thermalize the system to the desired temperature for at least $5 \times 10^{6} \mathrm{MC}$ sweeps. Then, we perform another $5 \times 10^{6} \mathrm{MC}$ sweeps, with measurements recorded every 500 sweeps. The spin configurations are then used as initial configurations (IC) for molecular dynamics (MD)\cite{zhang2019dynamical, moessner1998properties}, where each independent configuration is time-evolved deterministically according to the semi-classical Landau-Lifshitz-Gilbert equations of motion \cite{lakshmanan2011fascinating},
\begin{align}
\frac{d}{d t} \mathbf{S}_{i}=-\mathbf{S}_{i} \times \frac{\partial H}{\partial \mathbf{S}_{i}}.
\end{align}
We allow the system to evolve for a long but finite time of $t=300\  \mathrm{meV^{-1}}$, with step sizes of $\delta t=0.08\ \mathrm{meV^{-1}}$ to obtain $S_{i}^{\mu}(t) S_{j}^{\nu}(0)$. We then average over all the ICs to obtain $\left\langle S_{i}^{\mu}(t) S_{j}^{\nu}(0)\right\rangle$. These results are then numerically Fourier transformed to obtain the momentum- and energy-dependent dynamical structure factors, $\mathcal{S}(\mathbf{q}, \omega)$. Our classical results are lastly re-scaled by a factor of $\beta\omega$, where $\beta=1/k_BT$, in order to reflect the classical-quantum correspondence $\mathcal{S}_{\mathrm{classical}}(\mathbf{q}, \omega)=\beta\omega\mathcal{S}_{\mathrm{quantum}}(\mathbf{q}, \omega)$ in the linear spin-wave theory framework \cite{zhang2019dynamical}. 

\end{document}